\documentclass[11pt,paper,twoside]{JHEP3}

\usepackage{graphicx}                  
\usepackage{amsmath,amssymb,amsfonts}
\usepackage{xspace}                    
\usepackage{float}
\usepackage{dcolumn}
\usepackage{fancyhdr}
\usepackage{cite}
\usepackage{multirow}
\usepackage{booktabs}

\allowdisplaybreaks[1]

\def\words#1{\mbox{\small{\,#1}}}
\def\be{\begin{equation}}
\def\ee{\end{equation}}

\newcommand{\M}{\mathcal{M}}
\newcommand{\N}{\mathcal{N}}
\newcommand{\K}{\mathcal{K}}
\renewcommand{\vec}[1]{\mathbf{#1}}
\newcommand{\Order}{\mathcal{O}}
\newcommand{\order}{\mathcal{O}}
\newcommand{\mpn}{M_{\pi^0}}
\newcommand{\mpc}{M_\pi}
\newcommand{\meta}{M_{\eta}}
\newcommand{\tbk}{\hspace{-3.5mm}}
\newcommand{\tpm}{\tbk$\pm$}
\newcommand{\tr}{\text{tree}}
\newcommand{\CA}{\text{CA}}
\newcommand{\nn}{\nonumber\\}
\title{Rescattering effects in \boldmath{$\eta\to3\pi$} decays\thanks{Partial financial support by the Helmholtz Association through funds provided
to the virtual institute ``Spin and strong QCD'' (VH-VI-231), 
the project ``Study of Strongly Interacting Matter''
(HadronPhysics2, grant No.~227431) under the 7th Framework Programme of the EU, 
by DFG (SFB/TR 16, ``Subnuclear Structure of Matter'') 
and by the Bonn-Cologne Graduate School of Physics and Astronomy is gratefully
acknowledged.}}

\author{Sebastian P. Schneider,  Bastian Kubis, Christoph Ditsche\\
           Helmholtz-Institut f\"ur Strahlen- und Kernphysik (Theorie) and \\
           Bethe Center for Theoretical Physics,
           Universit\"at  Bonn\\
           Nu{\ss}allee 14--16, D-53115 Bonn, Germany \\
           E-mail: \email{schneider@hiskp.uni-bonn.de},~\email{kubis@hiskp.uni-bonn.de},\\
     \hspace{1.3cm}\email{ditsche@hiskp.uni-bonn.de}}

\abstract{
The isospin-breaking decay $\eta\to3\pi$ is an ideal tool to extract
information on light quark mass ratios from experiment.  For a
precise determination, however, a detailed description of the Dalitz
plot distribution is necessary.  In that respect, in particular the 
slope parameter $\alpha$ of the neutral decay channel causes some
concern, since the one-loop prediction from chiral perturbation theory
misses the experimental value substantially.  We use the modified non-relativistic effective field-theory,
a dedicated framework to analyze final-state interactions beyond one loop
including isospin-breaking corrections, to extract charged and neutral Dalitz plot
parameters. Matching to chiral perturbation theory at next-to-leading order, 
we find $\alpha = -0.025 \pm 0.005$, in marginal agreement with experimental findings.
We derive a relation between charged and neutral decay parameters that points towards a
significant tension between the most recent KLOE measurements of these observables.
}
\preprint{HISKP-TH-10/26}

\begin{document}

\section{Introduction}
\label{sec:intro}
\setcounter{footnote}{0}

The decay $\eta\to3 \pi$ has been the center of attention in many theoretical and experimental works over the recent decades. The considerable interest is due to the fact that the decay can only occur via 
isospin-breaking operators and is therefore sensitive to the up- and down-quark mass difference. Indeed, the $\eta\to 3\pi$ transition amplitude is inversely proportional to the quark mass double ratio $\mathcal{Q}^2$,
\be
\frac{1}{\mathcal{Q}^2}=\frac{m_d^2-m_u^2}{m_s^2-\hat{m}^2} ~, \qquad \hat{m}=\frac{1}{2}(m_u+m_d) ~, 
\ee
and thus the decay provides an excellent testing ground for the breaking of chiral symmetry. 

Despite valiant efforts it seemed difficult to bring theoretical description and experimental results in agreement. First attempts that relied on an electromagnetic transition~\cite{Sutherland,BellSuth} 
were unsuccessful in explaining the decay. SU(3) current algebra techniques in combination with the partially conserved axial-vector current hypothesis~\cite{Cronin,Osborn} were generalized to SU(3) 
chiral perturbation theory (ChPT) and initiated systematic improvements to the decay rate. While the one- and two-loop corrections to the decay were sizable~\cite{GLeta3pi,BGeta3pi}, a consistent implementation of 
electromagnetic contributions only lead to small effects~\cite{BKWeta3pi,DKMeta3pi}. Despite these theoretical improvements the Dalitz plot expansion, especially of the neutral decay, remained an unsolved puzzle. The slope $\alpha$ 
vanishes at leading order, while at next-to-leading order ($\Order(p^4)$, one loop) it disagrees in sign with experimental findings~\cite{CB@BNL,CB@LEAR,GAMS,KLOE,MAMI-B,MAMI-C,SND,WASA@CELSIUS,WASA@COSY}. 
The same holds for the next-to-next-to-leading order ($\Order(p^6)$, two loops) calculation~\cite{BGeta3pi}. 
The error on the final result is rather large, so that it allows for a negative slope parameter. 
However, this error is not based on the uncertainties due to the low-energy constants at $\order(p^6)$, 
which are estimated by resonance saturation, but results solely from the authors' fitting procedure. 

It has been argued that $\pi\pi$ final-state interactions are the dominant force behind the sizable corrections~\cite{Neveu,Roiesnel}, motivating several dispersive analyses~\cite{KWWeta3pi,ALeta3pi,CLPeta3pi,Zdrahaleta3pi} (see also Ref.~\cite{Gassereta3pi}), 
which were able to give a more robust prediction of the slope parameter. Among the shortcomings of these dispersion relation techniques and the next-to-next-to-leading-order calculation is the treatment of higher-order 
isospin-breaking effects due to electromagnetism, as for example the mass difference between charged and neutral pions. It is not yet clear how to incorporate these effects. 

An analysis of $\eta\to3\pi$ in the framework of unitarized chiral perturbation theory has been conducted in Ref.~\cite{BNeta3pi}, producing remarkable agreement with experiment. In particular, the experimental value of the
slope parameter in the neutral decay channel can be accommodated. However, since this approach is based on an elaborate fitting procedure, wherein the U(3) expansion parameters are determined from several hadronic 
$\eta$ and $\eta'$ decay channels, among those $\eta\to3\pi$, we do not consider this value for the slope parameter an unbiased \emph{prediction}.
Finally, a study of $\eta\to3\pi$ in the framework of 
resummed ChPT is currently work in progress~\cite{Kolesareta3pi}.

In this work we attempt to bridge the gap between the ChPT prediction and the dispersive analysis using the modified non-relativistic effective field theory framework (NREFT). While this framework
does not allow for a fundamental prediction of physical observables, it is ideally suited to study the dynamics of the final-state interactions. 
At two-loop accuracy and with the correct empirical $\pi\pi$ scattering parameters, 
we ought to have a reasonable approximation to the full dispersive resummation
of rescattering effects at hand, so when  
matching to ChPT at $\order(p^4)$, we can hope to find a transparent interpretation of the dispersive results
obtained in a similar fashion~\cite{KWWeta3pi}.
Additionally, the non-relativistic framework provides access to investigating the effects of isospin-breaking corrections. 

This article is organized as follows. In Sect.~\ref{sec:Dalexp} we begin with a short description of the Dalitz plot expansion 
and the conventions used throughout this work. An introduction to the non-relativistic framework, its power counting, the matching procedure, and numerical input is given in Sect.~\ref{sec:NREFT}. 
Sections~\ref{sec:isospinlimitresults} and~\ref{sec:isospinbreaking} comprise the analytic and numerical results in the isospin limit and with isospin breaking included. In 
Sect.~\ref{sec:RelatingDal} we study final-state interaction effects on an isospin relation between the charged and the neutral decay channel.
In Sect.~\ref{sec:widths} finally, we briefly comment on the $\eta\to3\pi$ partial widths and their ratio,
before summarizing our findings in Sect.~\ref{sec:summary}.
Several of the more laborious formulae are relegated to the appendices.

\section{Dalitz plot expansion of the decay amplitude}
\label{sec:Dalexp}
In the following we consider the charged and neutral decay modes 
\begin{align}
\eta(P_{\eta}) &\to \pi^+(p_1)\pi^-(p_2)\pi^0(p_3) ~,& s_1+s_2+s_3 &= 3s_c=\meta^2 + 2\mpc^2+\mpn^2~,\nn& & Q_c&=\meta-2\mpc-\mpn~,\nn
\eta(P_{\eta}) &\to \pi^0(p_1)\pi^0(p_2)\pi^0(p_3) ~,& s_1+s_2+s_3 &= 3s_n=\meta^2 + 3\mpn^2~,\nn& & Q_n&=\meta-3\mpn~,
\end{align}
where the kinematical variables are defined as $s_i=(P_{\eta}-p_i)^2$ with $p_i^2=M_i^2$, $i=1,2,3$, and $Q_{n/c}$ is the excess energy of the respective channel.\footnote{For convenience, we use a different notation from what is usually found in the literature. The transition can be made setting $s_1=t$, $s_2=u$, $s_3=s$.} We will use the notation $\mpc\doteq M_{\pi^{\pm}}$ throughout. 

In experimental analyses of these decays, the squared absolute value of the amplitude is conventionally 
expanded as a polynomial around the center of the Dalitz plot in terms of symmetrized coordinates. 
For the charged decay channel one uses
\be\label{eq:xydef}
 x=\sqrt{3}\frac{E_2-E_1}{Q_c}=\frac{s_1-s_2}{\sqrt{3}R_c}~, \qquad
 y=\frac{3E_3}{Q_c}-1=\frac{s_n-s_3}{R_c}+\delta~,
\ee
where $E_i=p_i^0-M_i~$ is the kinetic energy of $i$-th particle in the $\eta$ rest frame, and we used the definitions
\begin{equation}
 p_i^0=\frac{\meta^2+M_i^2-s_i}{2\meta}~,\qquad R_{c/n}=\frac{2}{3}\meta Q_{c/n}~, \qquad \delta=\frac{Q_n}{Q_c}-1~.
\end{equation}
For the neutral channel one defines
\begin{align}
\label{eq:zdef}
 &z=\frac{2}{3}\sum_{i=1}^{3}\Bigl(\frac{3E_i}{Q_n}-1\Bigr)^2=\frac{2}{3}\sum _{i=1}^3\frac{(s_i-s_n)^2}{R_n^2}=x_n^2+y_n^2~,\nn
 &\qquad\qquad x_n=\sqrt{z}\cos(\phi)~,\quad y_n=\sqrt{z}\sin(\phi)~,
\end{align}
where we have introduced polar coordinates in the center of the Dalitz plot.
These definitions of $x_n$ and $y_n$ agree with $x$ and $y$ only for $\mpc=\mpn$.
Experimental data is then fitted to the {\it Dalitz plot} distribution, which is  of the form
(assuming charge conjugation invariance)
\begin{align}\label{eq:ampsq}
 |\M_c(x,y)|^2&=|\N_c|^2 \big\{1+ay+by^2+dx^2+fy^3+gx^2y + \ldots \big\}~,\nn
 |\M_n(z)|^2&=|\N_n|^2 \big\{1+2\alpha z+2\beta z^{3/2}\sin(3\phi)+2\gamma z^2 + \ldots \big\}~,
\end{align}
where $a$, $b$, $d$, $f$, $g$ and $\alpha$, $\beta$, $\gamma$ are the Dalitz plot parameters and $\N_c$, $\N_n$ are the normalizations of the charged and the neutral decay, respectively. We note that of the 
higher-order parameters beyond quadratic order in $x$ and $y$, only $f$ has been measured so far (by the KLOE collaboration~\cite{KLOEcharged}). However, with the advent of very high statistics
measurements for $\eta\to3\pi^0$ e.g.\ at MAMI~\cite{Unverzagt}, a determination of $\beta$ and $\gamma$ might not be beyond the realm of possibility.

\FIGURE{
 \includegraphics[width=0.49\linewidth]{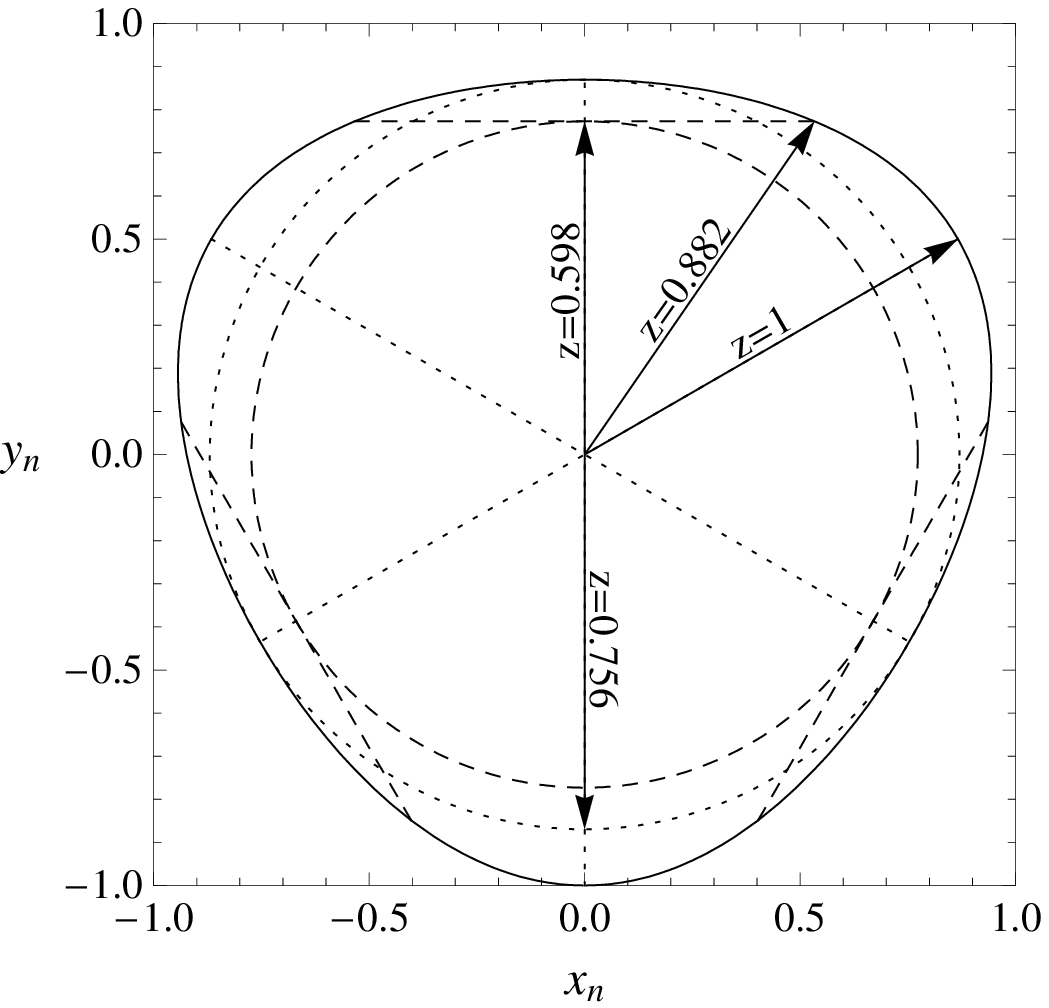}
 \caption{Boundary of the $\eta\to3\pi^0$ Dalitz plot. 
Dotted: symmetry axes and biggest enclosed circle. 
Dashed: cusps at $s_i=4\mpc^2$ and corresponding circle. 
Arrows: indicating specific $z$ values (see text for details).}
\label{fig:cusps}
}
\begin{sloppypar}
We wish to comment on the validity of the polynomial expansion Eq.~\eqref{eq:ampsq} in particular
for the neutral decay channel.
The boundary of the Dalitz plot for $\eta\to3\pi^0$ is shown in Fig.~\ref{fig:cusps}. The dotted lines denote the three symmetry axes, the dotted circle depicts the beginning of the rapid decrease of pure phase space for radii $\sqrt{z}>\sqrt{0.756}$. It is important to note that the cusps due to $\pi^+\pi^-\to\pi^0\pi^0$ final state rescattering occur at $s_i=4\mpc^2$ and not at a single $z$ value; the smallest and the largest values of $z$ crossing the cusp lines ($z=0.598$ and $z=0.882$, respectively) are indicated at the corresponding arrows. Therefore the polynomial representation for the neutral Dalitz plot distribution~\eqref{eq:ampsq} is only valid for $z<0.598$, i.e.\ inside the dashed circle.
\end{sloppypar}

Table~\ref{tab:DalExpTh} summarizes the latest experimental determinations and theoretical predictions for $\alpha$.
In the following we propose an explanation for the disagreement between the ChPT result and experimental data. 
Our findings substantiate the dispersive result~\cite{KWWeta3pi}, 
and we are confident that it leads to a better understanding of the nature of the final-state interactions. 

It is worthwhile at this point to quote the ChPT decay amplitudes at leading order $p^2$ 
and up to next-to-leading order in the isospin-breaking parameters $m_d-m_u$ and $e^2$.
For the charged and neutral decay, respectively, they read~\cite{DKMeta3pi}  
(we use the Condon--Shortley phase convention throughout)
\begin{align}\label{eq:LOChPTamps}
  \M^\textnormal{LO}_c(s_1,s_2,s_3)&=\frac{B_0(m_d-m_u)}{3\sqrt{3}F_\pi^2}\bigg\{1+\frac{3(s_3-s_n)}{\meta^2-\mpn^2}\bigg\}
~,\nn
  \M^\textnormal{LO}_n(s_1,s_2,s_3)&=-\frac{B_0(m_d-m_u)}{\sqrt{3}F_\pi^2} 
~, 
\end{align}
where $F_\pi=92.2$~MeV is the pion decay constant, and $B_0$ is linked to the quark condensate in the (SU(3)) chiral limit
in the standard manner. 
Equation~\eqref{eq:LOChPTamps} shows the isospin-violating nature of the decay, as both leading-order  amplitudes are explicitly of order $m_d-m_u$.
At that order in isospin breaking,
the $\eta\to3\pi$ amplitudes fulfill the well-known $\Delta I=1$ 
relation
\begin{align}\label{eq:isospinrel}
 \M_n(s_1,s_2,s_3)=-\M_c(s_1,s_2,s_3)-\M_c(s_2,s_3,s_1)-\M_c(s_3,s_1,s_2)~,
\end{align}
which can be easily checked in Eq.~\eqref{eq:LOChPTamps}.
This relation even holds in general at leading order in the isospin-breaking parameters,
i.e.\ also for terms of  $\Order(e^2)$~\cite{BKWeta3pi}, and
is only violated at $\Order((m_d-m_u)e^2$)~\cite{DKMeta3pi}.
In the following, we will often adopt a loose way of talking and 
speak about the \emph{isospin limit} for the charged and neutral $\eta\to 3\pi$ amplitudes;
this only refers to the approximation in which the relation Eq.~\eqref{eq:isospinrel} holds,
in particular $\mpn=\mpc$, 
and \emph{not} to the limit $m_u=m_d$, where the decay $\eta\to 3\pi$ is (almost) forbidden.

\TABLE[t]{
\centering
\renewcommand{\arraystretch}{1.2}
\begin{tabular}{c rcl}
\toprule
Theory & & $\alpha$						\\
\midrule
ChPT $\order(p^4)$~\cite{GLeta3pi} & $+0.013 $&				\\
ChPT $\order(p^6)$~\cite{BGeta3pi} & $+0.013 $&\tpm&\tbk$0.032$		\\
Dispersive~\cite{KWWeta3pi} & $-0.007$&\tbk$\ldots $&\tbk$-0.014$	\\
\midrule
Experiment & & $\alpha$						\\
\midrule
Crystal Ball@BNL~\cite{CB@BNL}& $ -0.031 $&\tpm&\tbk$0.004$	\\
Crystal Barrel@LEAR~\cite{CB@LEAR} & $ -0.052 $&\tpm&\tbk$0.020$\\
GAMS-2000~\cite{GAMS}& $ -0.022 $&\tpm&\tbk$0.023$	\\
KLOE~\cite{KLOE} & $ -0.0301$&\tpm&\tbk$0.0035_{-0.0035}^{+0.0022}$ \\
MAMI-B~\cite{MAMI-B} & $-0.032 $&\tpm&\tbk$0.002\pm0.002$	\\
MAMI-C~\cite{MAMI-C} & $-0.032 $&\tpm&\tbk$0.003$		\\
SND~\cite{SND} & $ -0.010 $&\tpm&\tbk$0.021\pm0.010$		\\
WASA@CELSIUS~\cite{WASA@CELSIUS} & $-0.026 $&\tpm&\tbk$0.010\pm0.010$	\\
WASA@COSY~\cite{WASA@COSY} & $-0.027 $&\tpm&\tbk$0.008\pm0.005$	\\
\bottomrule
\end{tabular}
\renewcommand{\arraystretch}{1.0}
\caption{Theoretical predictions and experimental findings on the slope parameter $\alpha$.\label{tab:DalExpTh}}}

Note furthermore that all contributions involving $\Delta_\pi = \mpc^2-\mpn^2 = \order(e^2)$
in the charged decay amplitude have been absorbed by writing Eq.~\eqref{eq:LOChPTamps} in terms of $s_n$.
This motivates an expansion of the decay amplitudes of \emph{both} channels around the point $s_3=s_n$, $s_1=s_2$:
we anticipate that, defined this way, higher-order isospin-breaking corrections to the $\Delta I =1$ rule for the
normalization of the amplitude are going to be of chiral order $p^4$,
without contributions from the tree-level amplitudes Eq.~\eqref{eq:LOChPTamps}, and therefore small.
This ``center'' of the Dalitz plot then corresponds to $s_1=s_2=s_3=s_n$ and $x_n=y_n=z=0$ in the neutral channel,
but to $s_1=s_2=s_n+\Delta_\pi$, $s_3=s_n$ or $x=0$ and $y=\delta\neq0$ in the charged case.
The charged and neutral decay amplitudes then take the form
\begin{align}\label{eq:ampsqS1S2S3}
  \M_c(s_1,s_2,s_3)&=\tilde\N_c\Big\{1+\tilde a(s_3-s_n)+\tilde b(s_3-s_n)^2+\tilde d(s_1-s_2)^2+\tilde f(s_3-s_n)^3\nn
		   &\qquad+\tilde g (s_1-s_2)^2(s_3-s_n)+ \ldots \Big\}\nn
		   &=\N_c\Big\{1+\bar a y+\bar b y^2 + \bar d x^2+\bar fy^3+\bar g x^2y+\ldots+\order\big((R_n-R_c)^2\big)\Big\}~,\nn
  \M_n(s_1,s_2,s_3)&=\N_n\Big\{1+\tilde\alpha \bigl[(s_1-s_n)^2+(s_2-s_n)^2+(s_3-s_n)^2\bigr]\nn
		   &\qquad+\tilde\beta\bigl[(s_1-s_n)^3+(s_2-s_n)^3+(s_3-s_n)^3\bigr]\nn
		   &\qquad+\tilde\gamma\bigl[(s_1-s_n)^4+(s_2-s_n)^4+(s_3-s_n)^4\bigr]+ \ldots\Big\}\nn
		   &=\N_n\Big\{1+\bar\alpha z+\bar\beta z^{3/2}\sin(3\phi)+\bar\gamma z^2+ \ldots \Big\}~.
\end{align}
The relations between the expansion parameters up to first order in isospin breaking are found to be
\begin{align}\label{eq:paramrel}
  \N_c&=\bar\N_c\times\tilde\N_c~,\qquad \bar\N_c=1+\tilde a (R_n-R_c)~,\nn
  \bar a &= -R_c\frac{\tilde a+2(R_n-R_c)\tilde b}{\bar \N_c}~,\quad
  \bar b = R_c^2\frac{\tilde b+3(R_n-R_c)\tilde f}{\bar \N_c}~,\quad 
  \bar d=3R_c^2\frac{\tilde d+\tilde g(R_n-R_c)}{\bar \N_c} ~,\nn
  \bar f &=-\frac{R_c^3\tilde f}{\bar \N_c}~,\quad
  \bar g=-\frac{3R_c^3\tilde g}{\bar \N_c}~,\quad
  \bar \alpha = \frac{3}{2}R_n^2\tilde\alpha~,\quad 
  \bar \beta=\frac{3}{4} R_n^3\tilde\beta~,\quad 
  \bar \gamma=\frac{9}{8}R_n^4\tilde\gamma~.
\end{align}
The expansion in powers of $R_n-R_c\simeq 3.35\times 10^{-3}$~GeV$^2$ hinges on the fact that we have considered isospin breaking corrections in the definition of $y$ (in the isospin limit, $R_n=R_c$, we reproduce the results derived
in Ref.~\cite{BGeta3pi}). The relations to the Dalitz plot parameters of the squared value of the respective amplitudes 
Eq.~\eqref{eq:ampsq} are then easily shown to be
\begin{align}\label{eq:paramfinal}
  a &= 2\,{\rm Re}(\bar a)~, &
  b &= |\bar a|^2+2\,{\rm Re}(\bar b)~,&
  d & = 2\,{\rm Re}(\bar d)~, &
  f &= 2\,{\rm Re}(\bar a\bar b^*+\bar f)~, \nn
  g &= 2\,{\rm Re}(\bar a\bar d^*+\bar g)~,&
\alpha &={\rm Re}(\bar \alpha)~,&
\beta &={\rm Re}(\bar \beta)~, &
\gamma&={\rm Re}(\bar \gamma)~.
\end{align}
The $\Delta I=1$ rule Eq.~\eqref{eq:isospinrel} gives rise to relations between Dalitz plot parameters and normalizations of the neutral and the charged decay amplitude, namely
\be
\label{eq:Dalrel}
  \N_n =-3\tilde \N_c~, \qquad
  \tilde\alpha = \frac{1}{3}(\tilde b+3\tilde d)~. 
\ee

\section{The modified non-relativistic effective field theory framework}
\label{sec:NREFT}
In this work we will use the modified non-relativistic effective field theory (NREFT) framework to analyze the 
final-state interactions in $\eta\to3\pi$. This framework provides a useful tool 
to investigate low-energy scattering and decay processes: it has found applications in detailed studies 
of cusp effects in $K\to3\pi$~\cite{K3pi,KL3pi,RadK3pi} and $\eta\to3\pi$~\cite{Gullstrom} 
as well as $\eta'\to\eta\pi\pi$~\cite{etapetapipi} decays, 
and has recently been extended to describe near-threshold pion photo- and electroproduction on the 
nucleon~\cite{pionphotoprod,pionelectroprod} 
(for an overview on cusp effects in meson decays, see Ref.~\cite{BKrev}). 

An analysis of $\eta\to 3\pi$ within the non-relativistic framework is useful for the following reasons. While the non-relativistic amplitude is perturbative, just as the chiral amplitude, it allows for a more accurate 
implementation of $\pi\pi$ interactions due to the inclusion of phenomenological threshold parameters
as determined from Roy equations. Non-perturbative treatments, as for example dispersive analyses, are expected 
to yield yet more precise results. Compared to such numerically very involved studies, however, the NREFT calculation leads to a very transparent analytic representation. Moreover, it allows for the direct implementation 
of isospin breaking in particular in all kinematic effects, which is much more involved in ChPT and unexplored in dispersive analyses. 

In that context it is useful to narrow down the precise definition of the term 
``non-relativistic'' as it is used in our work. Our representation of the decay amplitude is only non-relativistic in the sense that inelastic thresholds outside the physical region are subsumed into point-like effective coupling constants. Inside the physical region, 
however, we arrive at a fully covariant expression with the correct non-analytic low-energy behavior. The number of low-energy Dalitz plot couplings to be included in the Lagrangian at tree-level is modeled after 
the traditional (experimental) Dalitz plot expansion, which seems to yield a rather good description of the experimental data in the center of the Dalitz plot. We note again, see Sect.~\ref{sec:Dalexp}, that the \emph{full} Dalitz plot is not 
accurately described by a polynomial expansion, since such a representation neglects non-analytic effects, such as cusps at the opening of the charged pion threshold (see also Refs.~\cite{DKMeta3pi,KL3pi,Gullstrom}). 

In fact, the non-relativistic approach to $\eta\to3\pi$ is not new. In Ref.~\cite{Gullstrom} the authors performed a fit to experimental data in an attempt to investigate the cusp effect in $\eta\to 3\pi^0$ generated at the opening of 
the charged pion threshold. The scope of our work is entirely different. We focus specifically on an analysis of the Dalitz plot parameters based on numerical input parameters derived from ChPT. For that 
endeavor the amplitudes are calculated to yet-higher accuracy in order to ensure the incorporation of the most prominent effects generated by the final-state interactions. In the following section
we give a brief introduction to the modified non-relativistic framework.

\subsection{Power counting (1): basics and tree amplitudes}
\label{subsec:powcount}
A Lagrangian treatment of $\eta\to 3\pi$ in the non-relativistic framework is provided in Ref.~\cite{KL3pi} and will not be repeated here. Instead, we will briefly comment on the power counting and outline 
the basic features of the amplitudes of $\eta\to3\pi$ and the $\pi\pi$ final-state interactions. 

A consistent power counting scheme for the modified non-relativistic effective field theory is constructed by introducing the formal non-relativistic parameter $\epsilon$ and count
\begin{itemize}
\item pion 3-momenta (in the $\eta$ rest frame) as $\Order(\epsilon)$,
\item kinetic energies $T_i=p_i^0-M_i$ as $\order(\epsilon^2)$, 
\item masses of the particles involved as $\order(1)$, but $\Delta_\pi=\mpc^2-\mpn^2$ as $\order(\epsilon^2)$, 
\item and the excess energy $Q_{n/c}=\sum_i T_i$ as $\order(\epsilon^2)$. 
\end{itemize}
Loop corrections in the perturbative series involve $\pi\pi$ rescattering at not-too-high energies, which can be related to the effective range expansion of the $\pi\pi$ amplitude. Since these effective range parameters
are phenomenologically small, we use them as an additional power counting parameter, referred to generically as $a_{\pi\pi}$. We thus have a correlated expansion in $a_{\pi\pi}$ and $\epsilon$ and can uniquely 
assign powers to our loop expansion (for a more detailed introduction to the modified non-relativistic effective field theory we refer to Refs.~\cite{K3pi,RadK3pi,KL3pi}).

Following the previous counting scheme a Lagrangian framework can be constructed. From that the $\eta\to3\pi$ amplitude at tree level can be derived as
\begin{align}\label{eq:tree}
\M_n^\tr(s_1,&s_2,s_3)=K_0+K_1\Bigl[(p_1^0-\mpn)^2+(p_2^0-\mpn)^2+(p_3^0-\mpn)^2\Bigr]+\order(\epsilon^6)~,\nn
\M_c^\tr(s_1,&s_2,s_3)=L_0+L_1(p_3^0-\mpn)+L_2(p_3^0-\mpn)^2+L_3(p_1^0-p_2^0)^2+\order(\epsilon^6)~,
\end{align}
where the low-energy couplings $K_i,~L_i$ are of $\order(1)$ and are related to the traditional Dalitz plot, 
see Sect.~\ref{subsec:MatchingEta}.
The isospin relation Eq.~\eqref{eq:isospinrel} translates into
\be
\label{eq:NREFTsym}
K_0 = -(3L_0 + L_1 Q_n-L_3 Q^2_n )~, \qquad
K_1 = -(L_2 + 3L_3)~.
\ee
The number of constants included here corresponds to expanding the Dalitz plot
up to quadratic order; we briefly comment on the possible inclusion of cubic terms at tree level in Sect.~\ref{sec:numres}. 
We remark that the number of four independent tree-level couplings (in the isospin limit) chosen here
equals the number of subtraction constants in several of the dispersive analyses~\cite{ALeta3pi,CLPeta3pi}
(compare Ref.~\cite{Zdrahaleta3pi}, though).
Analogously, the $\pi\pi$ scattering amplitude can be determined. We consider the following final-state processes ($i$) ($\pi^a\pi^b\to\pi^c\pi^d$):
($00$) ($00;00$), ($x$) ($+-;00$), ($+0$) ($+0;+0$), and ($+-$) ($+-;+-$). 
Up to $\order(a_{\pi\pi}^2\epsilon^2)$ the threshold expansion of the amplitudes 
in the respective channels are given as
\begin{align}\label{eq:nonrelscatt}
 {\rm Re}\, T_{NR}^{00}&=2C_{00}+2D_{00}(s-s_{00}^{\rm thr})+2F_{00}(s-s_{00}^{\rm thr})^2+4C_x^2J_{+-}(s)+\ldots~,\nn
 {\rm Re}\, T_{NR}^{x}&=2C_{x}+2D_{x}(s-s_{x}^{\rm thr})+2F_{x}(s-s_{x}^{\rm thr})^2+\ldots~,\nn
 {\rm Re}\, T_{NR}^{+0}&=2C_{+0}+2D_{+0}(s-s_{+0}^{\rm thr})+2F_{+0}(s-s_{+0}^{\rm thr})^2-E_{+0}(t-u)+\ldots~,\nn
 {\rm Re}\, T_{NR}^{+-}&=2C_{+-}+2D_{+-}(s-s_{+-}^{\rm thr})+2F_{+-}(s-s_{+-}^{\rm thr})^2-E_{+-}(t-u)+\ldots~,
\end{align}
where $s_i^{\rm thr}$ denotes the threshold of the pertinent channel, $s_{00}^{\rm thr}=4\mpn^2$, $s_{x}^{\rm thr}=4\mpc^2$, $s_{+0}^{\rm thr}=(\mpn+\mpc)^2$, $s_{+-}^{\rm thr}=4\mpc^2$. The one-loop function 
of the non-relativistic theory,
\begin{align}\label{eq:loopJ}
 J_{+-}(s)=\frac{i}{16\pi}\sqrt{1-\frac{4\mpc^2}{s}}~,
\end{align}
is responsible for a cusp structure in the ($00$)-channel (see Refs.~\cite{K3pi,MMS} for further details). The low-energy couplings are matched to the effective range expansion in the following section.

\subsection[Matching (1): $\pi\pi$ scattering]{Matching (1): \boldmath{$\pi\pi$} scattering}
\label{subsec:MatchingPipi}
We want to make more sense of the low-energy couplings introduced in the previous section. 
To determine the matching relations for the low-energy constants of $\pi\pi$ scattering, we resort to the effective range expansion of the $\pi\pi$ scattering amplitude, which is conventionally decomposed into 
partial waves according to
\begin{align}
 T_I(s,t)=32\pi \sum_l(2l+1)t_l^I(s)P_l(z)~,
\end{align}
where $t_l^I(s)$ is the partial wave amplitude of angular momentum $l$ and isospin $I$, $P_l(z)$ are the Legendre polynomials, and $z=\cos\theta$ is the cosine of the scattering angle in the center-of-mass system. Close to threshold the partial wave 
amplitude can be expanded in terms of the center-of-mass momentum $q^2\doteq q^2(s)=(s-4\mpc^2)/4$, leading to
\begin{align}
 {\rm Re}\,t_l^I(s)=q^{2l}\{a_l^I+b_l^Iq^2+c_l^Iq^4+\order(q^6)\}~,
\end{align}
where $a_l^I$ is the scattering length, $b_l^I$ is the effective range, and $c_l^I$ is the (leading) shape parameter. In the following we use the 
simplified notation $a_I,b_I,c_I$, as only S- and P-waves will be considered. 
In the language of NREFT power counting the previous equation is an expansion in orders of $\epsilon$, 
since $q^2\propto\epsilon^2$. The effective range expansion is thus
naturally related to the non-relativistic $\pi\pi$ scattering amplitude in Eq.~\eqref{eq:nonrelscatt}, 
and we can read off the matching relations for the low-energy couplings, shown here for simplicity in the isospin limit:
\begin{align}\label{eq:pipiMatch}
C_{00} &= \frac{16\pi}{3} (a_0+2a_2)~,& D_{00} &= \frac{4\pi}{3}(b_0+2b_2)~,& F_{00}&=\frac{\pi}{3}(c_0+2c_2)~, \\
C_x    &= \frac{16\pi}{3} (-a_0+a_2)~,& D_x &= \frac{4\pi}{3}(-b_0+b_2)~,& F_x &=\frac{\pi}{3}(-c_0+c_2)~,\nn
C_{+0} &= 8\pi a_2~,& D_{+0} &= 2\pi b_2 ~,& F_{+0}  &=\frac{\pi}{2} c_2~,& E_{+0} &= 12\pi a_1~,\nn
C_{+-} &= \frac{8\pi}{3} (2a_0+a_2)~,& D_{+-} &= \frac{2\pi}{3}(2b_0+b_2)~,& F_{+-} &= \frac{\pi}{6} (2c_0+c_2)~, & E_{+-}  &= 12\pi a_1~.\nonumber
\end{align}
Isospin-breaking corrections to these matching relations are discussed in Appendix~\ref{app:EMcorr}.
Note that Eq.~\eqref{eq:pipiMatch} is only valid up to $\Order(a_{\pi\pi}^2)$, i.e.\ 
$\pi\pi$ scattering to one loop, or $\eta\to3\pi$ to two loops.
At higher loop orders, the low-energy couplings $D_i$ and $F_i$ are renormalized,
which we will briefly discuss in the context of higher-loop resummation at the end of
Sect.~\ref{subsec:isolim}.

\begin{sloppypar}
We will use two sets of phenomenological values for the $\pi\pi$ effective range parameters,
the combined Roy equation plus ChPT analysis of Refs.~\cite{ACGL,CGL} (henceforth denoted by ACGL) and a combination of forward dispersion relations 
and Roy equations~\cite{Yndurain} (KPY). 
The central or ``best'' values for S- and P-wave scattering lengths and effective ranges 
as obtained in those two analyses are quoted in Table~\ref{tab:effrangeparam}.
The determination of the shape parameters is a little more delicate. 
We use the respective parameterizations of the phase shifts given in Refs.~\cite{ACGL,Yndurain} 
and calculate the scattering amplitude according to
\be
{\rm Re}\, t_0^I(q^2)=\Bigl(1+\frac{\mpc^2}{q^2}\Bigr)^{1/2}\frac{\tan\delta_I}{1+\tan^2\delta_I}~,\qquad I=0,\,2~.
\ee
Since the shape parameters are numerically very small in comparison to effective ranges and scattering lengths, 
they are rather sensitive to the method by which they are determined. 
\TABLE{
\renewcommand{\arraystretch}{1.2}
\begin{tabular}{l r r}
\toprule
			& \text{ACGL} 	& \text{KPY}	\\
\midrule
$a_0$			& $ 0.220$ 	& $0.223$	\\
$a_2$			& $ -0.0444$   	& $-0.0444$	\\
$b_0\times \mpc^2$	& $ 0.276$   	& $0.290$	\\
$b_2\times \mpc^2$	& $ -0.0803$   	& $-0.081$	\\
$c_0\times 10^2\mpc^4$	& $ -0.19$ 	& $0.04$	\\
$c_2\times 10^2\mpc^4$	& $ 1.33$	& $0.68$	\\
$a_1\times 10\mpc^2$	& $ 0.379$	& $0.381$	\\
$b_1\times 10^2\mpc^4$	& $ 0.567$	& $0.512$	\\
\bottomrule
\end{tabular}
\renewcommand{\arraystretch}{1.0}
\caption{Input values for the scattering lengths $a_I$, effective ranges $b_I$, and shape parameters $c_I$ as determined from the two parameterizations ACGL~\cite{ACGL,CGL} and KPY~\cite{Yndurain} (see text for discussion).\label{tab:effrangeparam}}}
\noindent For example, one
receives rather different results when extracting the shape parameter from a strict threshold expansion of the amplitude, 
or from a fit over a certain low-energy range,
minimizing the $\chi^2$-function
\be\label{eq:chi2}
 \chi^2(c_I)=\Bigl({\rm Re}\, t_0^I(q^2)-a_I-b_Iq^2-c_Iq^4\Bigr)^2~,
\ee
in a range from the threshold $4\mpc^2$ up to the expansion point $s_n$. 
Furthermore, the inclusion of an additional term $d_Iq^6$ causes significant deviations in the $I=0$ channel, 
since this term and the leading shape parameter are of comparable size. 
We decide to use the central values obtained from the minimization of Eq.~\eqref{eq:chi2} 
as the most reasonable approximation to the true partial wave.
The numerical results for $c_{0,2}$ thus obtained are also given in Table~\ref{tab:effrangeparam}.
In the following, we use the variation between the central values of the two parameterizations~\cite{ACGL,Yndurain} 
as a means to estimate the uncertainty due to $\pi\pi$ rescattering.
\end{sloppypar}

\subsection[Matching (2): $\eta\to3\pi$]{Matching (2): \boldmath{$\eta\to3\pi$}}
\label{subsec:MatchingEta}
We compare Eqs.~\eqref{eq:ampsqS1S2S3} and~\eqref{eq:tree} to derive 
the matching relation between the low-energy couplings of the $\eta\to3\pi$ tree amplitude and the traditional Dalitz plot parameterization, namely
\begin{align}\label{eq:matchtree}
K_0 &= \tilde\N_n^\tr\left(1-3 \tilde\alpha^\tr R_n^2\right)~,& K_1 &= 4\tilde\N_n^\tr\meta^2\tilde\alpha^\tr,\nn
L_0 &= \tilde\N_c^\tr(1+\tilde a^\tr R_n+\tilde b^\tr  R_n^2)~,& L_1 &= -2 \tilde\N_c^\tr \meta (\tilde a^\tr + 2\tilde b^\tr  R_n)~,\nn 
L_2 &= 4\tilde\N_c^\tr\meta^2\tilde b^\tr ~,& L_3 &= 4\tilde\N_c^\tr\meta^2\tilde{d}^\tr~,
\end{align}
where the superscript ``tree'' denotes tree-level input parameters. 
Note that Eq.~\eqref{eq:matchtree} fulfills the isospin relation
Eq.~\eqref{eq:NREFTsym} as long as Eq.~\eqref{eq:Dalrel} is satisfied. 
To extract the Dalitz plot parameters in the non-relativistic framework, we have to fix the numerical input for the tree-level low-energy couplings for the $\eta\to3\pi$ amplitude.
We determine the low-energy couplings of the Dalitz plot in Eq.~\eqref{eq:matchtree} by matching the non-relativistic framework to the one-loop ChPT amplitude~\cite{GLeta3pi} at the center of the Dalitz plot. 
Following Ref.~\cite{DKMeta3pi}, we evaluate the chiral $\eta\to3\pi$ amplitude
using \emph{neutral} masses everywhere.

We remark that the upcoming~\cite{CLPeta3pi} (and previous~\cite{ALeta3pi}) dispersive analyses use the Adler zero 
of the $\eta\to\pi^+\pi^-\pi^0$ amplitude as the matching point, compare Eq.~\eqref{eq:LOChPTamps}. 
It is protected by SU(2) symmetry and therefore not prone to large strange-quark-mass corrections. 
The chiral series is thus expected to converge rather quickly, which makes the Adler zero a natural choice. 
The fact that it lies outside the physical region at roughly $s_A\approx\frac{4}{3}\mpc^2$, 
however, renders matching the non-relativistic framework 
to the chiral amplitude at this point ill-fated: the expansion in terms of $\epsilon$ does not necessarily converge there,
and we therefore have to resort to matching inside the Dalitz plot.

For the matching procedure we tune the rescattering parameters in the non-relativistic amplitude in such a way as to mimic the chiral amplitude. In essence this means that the scattering lengths and effective
ranges are fixed at their current algebra values (this corresponds to the insertion of $\order(p^2)$ vertices in the chiral expansion). Explicitly, we have
\begin{align}\label{eq:CApipi}
  a_0^\CA &= \frac{7\mpc^2}{32\pi  F_\pi^2~}~,& a_2^\CA &= -\frac{\mpc^2}{16\pi F_\pi^2}~, &a_1^\CA &= \frac{1}{24\pi F_\pi^2}~,\nn
  b_0^\CA &= \frac{1}{4\pi F_\pi^2}~,& b_2^\CA &= -\frac{1}{8\pi F_\pi^2}~.
\end{align}
We proceed analogously with the $\eta\to3\pi$ couplings that enter the non-relativistic amplitude at one-loop level and derive from Eq.~\eqref{eq:LOChPTamps}
\begin{align}
  \tilde \N_c^\textnormal{LO} &= -\frac{(\meta^2 - \mpc^2)(\mpc^2+3\meta^2)}{16\mathcal{Q}^2 \sqrt{3} F_\pi^2 \mpc^2}~,\qquad \tilde{a}^\textnormal{LO}=\frac{3}{\meta^2-\mpc^2}~.
\end{align}
For our numerical analysis we will use the value for $\mathcal{Q}$ dictated by Dashen's theorem, $\mathcal{Q}_D=24.2$. 
Note that the specific choice does not hold any ramifications for our main statements, since it merely enters in the normalization, which drops out in the Dalitz plot parameters. 

The above matching procedure is consistent as it ensures that the imaginary parts are exclusively generated by $\pi\pi$ final-state interactions. Residual effects from the chiral pion loops are purely real
and  absorbed in the low-energy couplings. We use matching to $\order(p^4)$ and not to $\order(p^6)$ for practical reasons:
the above matching procedure is simpler and our results can be used to compare with and
interpret the dispersive analyses directly. 
A high-precision determination of the Dalitz plot parameters would likely require matching to $\order(p^6)$, 
but for that purpose the low-energy constants showing up at $\order(p^6)$ may 
not be known with sufficient accuracy. Numerically we obtain from matching to the ChPT amplitude at $\order(p^4)$ 
(using the chiral SU(3) low-energy constant\footnote{The effects of varying $L_3^{\rm r}$ within its error were checked
to be tiny compared to other uncertainties. We therefore only use the central value.} $L_3^{\rm r}=-3.5\times10^{-3}$~\cite{L3})
\begin{align}\label{eq:input}
  \tilde{\N}_c^\tr &= -0.158~,  & \tilde{a}^\tr &= 13.428~{\rm GeV}^{-2}~,\nn 
  \tilde{b}^\tr &= -7.291~{\rm GeV}^{-4}~,  & \tilde{d}^\tr &= 5.189~{\rm GeV}^{-4}~.
\end{align}

The particle masses used throughout this work are given by the current particle data group values~\cite{pdg}, i.e.\ $\mpc = 139.57$~MeV, $\mpn = 134.98$~MeV, and $\meta= 547.86$~MeV.

\subsection[Power counting (2): loops and $\eta\to3\pi$]{Power counting (2): loops and \boldmath{$\eta\to3\pi$}}
\label{subsec:powcounteta3pi}
The power counting scheme discussed in Sect.~\ref{subsec:powcount} gives rise to a natural decomposition of the NREFT amplitude. This can be seen as follows. The modified non-relativistic propagator
counts as $\order(\epsilon^{-2})$ (see e.g.\ Ref.~\cite{K3pi}), 
the loop integration measure (with one energy and three momentum integration variables) as $\epsilon^5$,
therefore any loop integral with two-body rescattering contributes at $\Order(\epsilon)$. 
Moreover, such a loop always involves a $\pi\pi$ rescattering vertex
and is thus of $\order(a_{\pi\pi})$. The decomposition of the full $\eta\to3\pi$ amplitude according to its loop-structure,
\be
\label{eq:full}
\M_{n/c}(s_1,s_2,s_3)= \M_{n/c}^\tr(s_1,s_2,s_3)+\M_{n/c}^{\text{1-loop}}(s_1,s_2,s_3)+\M_{n/c}^{\text{2-loop}}(s_1,s_2,s_3)+\ldots~,
\ee
is thus an expansion in powers of $a_{\pi\pi}\epsilon$. There is an interesting simplification of Eq.~\eqref{eq:full} close to the center of the Dalitz plot ($s_1\approx s_2$, $s_3\approx s_n$) above all two-pion thresholds. The contribution of
the one-loop function is purely imaginary as can be seen from Eq.~\eqref{eq:loopJ}. At the same time the two-loop bubble diagram, which is the product of two one-loop functions, is purely real and it can be shown that the imaginary part of the 
non-trivial two-loop function does not contribute at this order (see for example Refs.~\cite{K3pi,KL3pi}). Symbolically we can write both amplitudes in terms of the power counting parameter $a_{\pi\pi}$,
\begin{align}
  \M = \M_\tr + i \M_{\text{1-loop}} a_{\pi\pi} + \M_{\text{2-loop}} a_{\pi\pi}^2+\order(i a_{\pi\pi}^3\epsilon^3,ia_{\pi\pi}^2\epsilon^4)~,
\end{align}
where the $\order(ia_{\pi\pi}^2\epsilon^4)$ term stems from the three-particle cut at two-loop order, which is numerically small as discussed in Appendix~\ref{app:3cut} and therefore neglected. By taking the absolute value 
squared we obtain
\begin{align}\label{eq:symbexp}
  |\M|^2 = \M_\tr^2 + (\M_{\text{1-loop}}^2 +\M_\tr\times\M_{\text{2-loop}}) a_{\pi\pi}^2+\order(a_{\pi\pi}^4\epsilon^4,a_{\pi\pi}^3\epsilon^5)~.
\end{align}
We therefore expect one- and two-loop effects to be of the same size at the center of the Dalitz plot,
as only the two-loop contributions can interfere with the dominant tree terms there, 
and thus to impact the Dalitz plot parameters about equally. 

The heightened importance of rescattering effects in Dalitz plot parameters is further substantiated by another observation. Consider the generic one-loop function of $\pi\pi$ rescattering in 
the non-relativistic theory expanded about the center of the Dalitz plot ($s=s_n$, we neglect isospin-breaking effects in the following discussion, so that $\mpn=\mpc$):
\begin{align}
 J(s)&=\frac{i \sqrt{1-\frac{4 \mpc^2}{s_n}}}{{16\pi}}\Bigl(1+\frac{6\mpc^2 }{s_n}\frac{s-s_n}{\meta^2-9\mpc^2}-\frac{18\mpc^2(s_n -3 \mpc^2)}{s_n^2}\Bigl(\frac{s-s_n}{\meta^2-9\mpc^2}\Bigr)^2+\ldots\Bigr)\nn
     &=\order(\epsilon)~,
\end{align}
since $s-s_n=\order(\epsilon^2)$ and $\meta-3\mpc=\order(\epsilon^2)$. The same holds true for the two-loop functions. This implies that contributions to higher-order Dalitz plot parameters from the loop functions are 
enhanced non-analytically in $\meta-3\mpc$. We conclude from Eq.~\eqref{eq:symbexp}
\begin{align}
\M_{\text{1-loop}}^2 +\M_\tr\times\M_{\text{2-loop}}=\order(\epsilon^2)~,
\end{align}
which has substantial consequences for the slope parameter $\alpha$ of the neutral decay channel. 
We can parameterize the slope parameter according to
\begin{align}\label{eq:alphaloopexp}
  \alpha = \alpha_0 + \alpha_2 a_{\pi\pi}^2 + \order(a_{\pi\pi}^4)~.
\end{align}
From relations Eq.~\eqref{eq:matchtree} we find $\tilde\alpha^\tr=\order(1)$ and consequently the slope parameter at tree-level is of order 
$\alpha_0\propto Q_n^2\tilde\alpha^\tr=\order(\epsilon^4)$, whereas rescattering effects enter the slope parameter at $\order(a_{\pi\pi}^2\epsilon^2)$. This obviously implies that rescattering effects become 
increasingly more important for higher-order Dalitz plot parameters. 
On the other hand, they are far less significant (as we will confirm numerically below) for the 
normalization of the amplitude, for which we expect higher-order quark-mass renormalization effects to be more important.

\FIGURE[t]{
\centering
\includegraphics[width=0.9\linewidth]{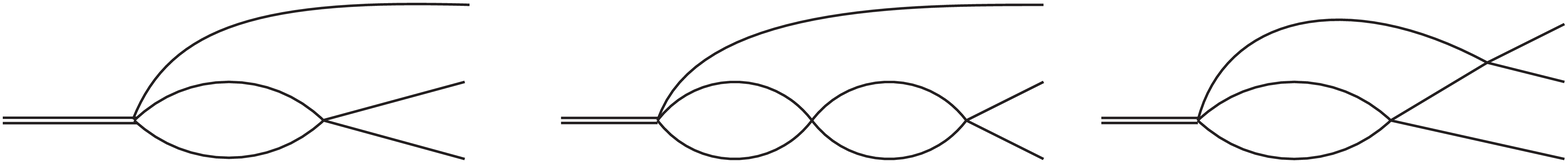}
\caption{The Feynman graph topologies at one and two loops contributing to the decay $\eta\to3\pi$ 
in NREFT.  The double line denotes the $\eta$ particle, the single lines stand for pions 
(of arbitrary charges).}
\label{fig:twolooptopologies}
}
The full NREFT representation beyond tree level with isospin breaking included is given in Appendix~\ref{app:NREFTrep}. 
It comprises the loop graph topologies displayed in Fig.~\ref{fig:twolooptopologies},
and is fully consistent in terms of non-relativistic power counting
up-to-and-including $\order(a_{\pi\pi}^2\epsilon^4)$, i.e.\ the vertices of the two-loop graphs
are included at $\order(\epsilon^2)$. 
Phenomenologically, one finds that the expansions of the $\eta\to3\pi$ and $\pi\pi\to\pi\pi$ 
polynomials in powers of $\epsilon^2$ only converge well starting from next-to-next-to-leading order,
i.e.\ the $\Order(\epsilon^2)$ terms (the linear slope in $\eta\to\pi^+\pi^-\pi^0$ and the 
$\pi\pi$ effective ranges) are not really suppressed compared to the leading (constant) terms.
This observation is readily understood resorting to chiral perturbation theory:
due to the Goldstone nature of the pions, the constant terms are chirally suppressed
by powers of $M_\pi^2$ and the leading $\Order(p^2)$ amplitudes are linear in energy $s$.
In other words, the $\Order(\epsilon^2)$ contributions are ``suppressed'' versus the constant ones
by factors of $s/M_\pi^2$, and only starting from $\Order(\epsilon^4)$, the relative suppression
is $s/\Lambda_\chi^2$ with $\Lambda_\chi \approx 1$~GeV.  
The chiral two-loop or $\Order(p^6)$ calculation~\cite{BGeta3pi} contains all the leading $\Order(p^2)$
vertices and therefore the linear $\eta\to3\pi$ slope as well as effective ranges for the $\pi\pi$
interaction (although not quite the phenomenologically accurate ones).
In order to guarantee that our NREFT representation of the decay amplitude is at least as accurate 
as the chiral two-loop one, we include all combinations of linear energy dependences in 
the three vertices of the two-loop diagrams.
Thus, our amplitude also contains terms that are of $\order(a_{\pi\pi}^2\epsilon^6)$ and $\order(a_{\pi\pi}^2\epsilon^8)$,
and due to the enhancement discussed above, the numerically most important ones
appearing at those orders.
The representation of the ``double bubbles'' (see also Fig.~\ref{fig:twolooptopologies}) is even strictly complete up to $\Order(a_{\pi\pi}^2\epsilon^6)$, as
P-wave contributions only start at $\Order(a_{\pi\pi}^2\epsilon^8)$. 
Furthermore, we have added shape parameter terms in the ``double bubbles'' and in the outer vertex of the 
irreducible two-loop graph, where the addition of these terms is trivial.

\section{The isospin limit}\label{sec:isospinlimitresults}
\begin{sloppypar}
We first give an analytic and numerical treatment of the amplitude in the isospin limit, which we \emph{define} as $\mpn=\mpc$ and using Eq.~\eqref{eq:NREFTsym}. 
This already includes the gross features of our total analysis. However, in the isospin limit, we can give relatively 
simple closed analytic expressions for all parts of the amplitudes up to two loops.
\end{sloppypar}
\subsection{Structure of the amplitude}\label{subsec:isolim}
\begin{sloppypar}
The non-relativistic decay amplitude (for the charged channel) can be split into parts consisting of tree and final-state contributions
\be\label{eq:split}
 \M_c(s_1,s_2,s_3) =\M_c^\tr(s_1,s_2,s_3)+\M_c^{\text{fsi}}(s_1,s_2,s_3)~,
\ee
where the tree amplitude is given by
\begin{align}
 \M_c^{\text{tree}}(s_1,s_2,s_3)=\tilde\N_c^\tr\Bigl\{1+\tilde a^\tr (s_3-s_n)+\tilde b^\tr(s_3-s_n)^2+\tilde d^\tr(s_1-s_2)^2\Bigr\}~,
\end{align}
and the rescattering contributions of the amplitude can be decomposed up to $\order(p^8)$ according to the isospin structure of the final-state pions~\cite{ALeta3pi,Stern,Knecht},
\begin{align}\label{isospindecomp}
\M_c^{\text{fsi}}(s_1,s_2,s_3)=\M_0(s_3)+(s_3-s_1)\M_1(s_2)+(s_3-s_2)\M_1(s_1)\nn+\M_2(s_1)+\M_2(s_2)-\frac{2}{3}\M_2(s_3)~,
\end{align}
where the index $I=0,1,2$ of the function $M_I(s_i)$ denotes the total isospin of the respective kinematic channel. At $\order(a_{\pi\pi}^2\epsilon^4)$ (for details see Sect.~\ref{subsec:powcounteta3pi}) the isospin amplitudes are given as
\begin{align}\label{eq:isoamps}
\M_0(s) &= \frac{5}{3}\biggl\{\ell_0(s)J(s)\Bigl(1+16\pi  a_0(s)J(s)\Bigr)\nn
& + \frac{32\pi}{3}\biggl[\Bigl(\ell_0'(s) a_0(\tilde{s})+2\ell_2'(s) a_2(\tilde{s})\Bigr)F^{(0)}(s) 
+ \biggl(\frac{2L_1}{\meta}\Bigl(\frac{2}{5} a_0(\tilde{s})-a_2(\tilde{s})\Bigr) - \ell_0'(s)b_0\nn
&\qquad-2\ell_2'(s) b_2\biggr)\frac{\meta{\mathbf{Q}}^2}{2Q^0}F^{(1)}(s) -2L_1\Bigl(\frac{2}{5}b_0-b_2\Bigr)\frac{\meta{\mathbf Q}^4}{4{Q^0}^2}F^{(2)}(s)\biggr]\biggr\}
16\pi  a_0(s) ~,\nn
\M_1(s) &=\biggl\{-\frac{q^2\ell_1(s)}{\meta}J(s) + \frac{80\pi s}{\meta Q^0}\biggr[\Bigl(\ell_0'(s) a_0(\tilde{s})-\ell_2'(s)a_2(\tilde{s})\Bigr)\big(F^{(0)}(s)-2F^{(1)}(s)\big)\nn
&\qquad+\biggl(\frac{L_1}{\meta}\Bigl(\frac{4}{5} a_0(\tilde{s})+ a_2(\tilde{s})\Bigr)-\ell_0'(s)b_0+\ell_2'(s)b_2\biggr)\frac{\meta{\mathbf{Q}}^2}{2Q^0}\big(F^{(1)}(s)-2 F^{(2)}(s)\big)\nn
&\qquad-L_1\Bigl(\frac{4}{5}b_0+ b_2\Bigr)\frac{\meta{\mathbf{Q}}^4}{4{Q^0}^2}\big(F^{(2)}(s)-2F^{(3)}(s)\big)\biggr]\biggr\}4\pi a_1(s) ~,\nn
\M_2(s) &= \biggl\{\ell_2(s)J(s)\big(1+16\pi a_2(s) J(s)\big)\nn
&+\frac{16\pi}{3}\biggl[\Bigl(5\ell_0'(s)a_0(\tilde{s})+\ell_2'(s)a_2(\tilde{s})\Bigr)F^{(0)}(s)+\biggl(\frac{4L_1}{\meta}\Bigl(a_0(\tilde{s})-\frac{a_2(\tilde{s})}{4}\Bigr)-5\ell_0'(s) b_0\nn
&\qquad-\ell_2'(s)b_2\biggr)\frac{\meta{\mathbf{Q}}^2}{2Q^0}F^{(1)}(s)-4L_1\Bigl(b_0-\frac{b_2}{4}\Bigr)\frac{\meta{\mathbf{Q}}^4}{4{Q^0}^2}F^{(2)}(s)\biggr]\biggr\}16\pi a_2(s) ~,
\end{align}
where the various polynomials are given by
\begin{align}
\ell_0(s)&=\frac{3}{5}\ell(s)+\frac{2}{5}\ell_2 (s)~,\quad 
\ell(s) = L_0 + L_1\left(p^0-\mpc\right)+L_2\left(p^0-\mpc\right)^2+L_3\frac{4\vec{Q}^2}{3s}\,q^2 ~,\nn
\ell_2(s) &= L_0 + L_1\Bigl(\frac{Q^0}{2}-\mpc\Bigr)
+ L_2\biggl[\Bigl(\frac{Q^0}{2}-\mpc\Bigr)^2+\frac{\vec{Q}^2}{3s}\,q^2\biggr]+ L_3\biggl[\Bigl(\frac{Q^0}{2}-p^0\Bigr)^2+\frac{\vec{Q}^2}{3s}\,q^2\biggr] ~,\nn
\ell_1(s) & =L_1+2L_2\Bigl(\frac{Q^0}{2}-\mpc\Bigr)+2L_3\Bigl(p^0-\frac{Q^0}{2}\Bigr) ~,\quad
\ell_0'(s)=\frac{3}{5}\ell'(s)+\frac{2}{5}\ell_2'(s)~, \nn
\ell'(s) &= L_0 + L_1 \Big( \frac{s}{2Q^0} -\mpc \Big)~,  \quad
\ell_2'(s) = L_0 + L_1 \Big( \frac{\meta}{2}- \mpc - \frac{s}{4Q^0}\Big)~, \nn
 a_I(s) & =a_I+b_Iq^2+c_Iq^4~,
\end{align}
and we use the kinematic variables
\begin{align}
 p^0&=\frac{\meta^2+\mpc^2-s}{2\meta}~,~~
Q^0 = \frac{\meta^2-\mpc^2+s}{2\meta} ~,~~
\mathbf{Q}^2 = \frac{\lambda(\meta^2,\mpc^2,s)}{4\meta^2} ~,\nn
\tilde s&= 2\mpc^2-s+\frac{\meta}{Q^0}\left(s+2\mathbf{Q}^2\right) ~,
\end{align}
with the K\"all\'en function $\lambda(x,y,z)=x^2+y^2+z^2-2(xy+yz+zx)$.
Note that the shape parameter terms $\propto c_I$ are to be omitted in $a_I(\tilde s)$;
we also neglect them in the $I=1$ partial wave.
We use the shorthand expressions $J(s)\doteq J_{+-}(s)$ and 
$F^{(n)}(s) \doteq F_+^{(n)}(\mpc,\mpc,\mpc,\mpc,s)$ (in the isospin limit), where for the exact form of the two-loop functions
we refer to Appendix~\ref{app:NREFTrep}. 
We can now write the Dalitz plot parameters in terms of the isospin amplitudes, namely for the charged channel
\begin{align}\label{eq:chargeddef}
 \N_c&=\tilde\N_c^\tr+\M_0(s_n)+\frac{4}{3}\M_2(s_n)~,\nn
 \bar a&= - \frac{R_c}{\N_c}\Bigl(\tilde\N_c^\tr \tilde a^\tr + \M_0^{(1)}(s_n)+3\M_1(s_n)-\frac{5}{3}\M_2^{(1)}(s_n)\Bigr)~,\nn
 \bar b&= \frac{R_c^2}{\N_c} \Bigl(\tilde\N_c^\tr \tilde b^\tr+\frac{1}{2}\M_0^{(2)}(s_n)-\frac{3}{2} \M_1^{(1)}(s_n)-\frac{1}{12}\M_2^{(2)}(s_n)\Bigr)~,\nn
 \bar d&=\frac{3R_c^2}{\N_c} \Bigl(\tilde\N_c^\tr \tilde d^\tr + \frac{1}{2}\M_1^{(1)}(s_n)+\frac{1}{4}\M_2^{(2)}(s_n)\Bigr)~,\nn
 \bar f&=-\frac{R_c^3}{\N_c} \Bigl(\frac{1}{6}\M_0^{(3)}(s_n)+\frac{3}{8} \M_1^{(2)}(s_n)-\frac{11}{72}\M_2^{(3)}(s_n)\Bigr)~,\nn
 \bar g&=-\frac{3R_c^3}{8\N_c}\Bigl(\M_1^{(2)}(s_n)- \M_2^{(3)}(s_n)\Bigr)~,
\end{align}
and for the neutral channel
\begin{align}\label{eq:neutraldef}
\bar \alpha&=\frac{R_n^2}{4\N_c} \Bigl(2\tilde\N_c^\tr\big(\tilde b^\tr+3\tilde d^\tr\big)+\M_0^{(2)}(s_n)+\frac{4}{3} \M_2^{(2)}(s_n)\Bigr)~,\nn
\bar \beta&=\frac{R_n^3}{24\N_c}\Bigl( \M_0^{(3)}(s_n)+\frac{4}{3} \M_2^{(3)}(s_n)\Bigr)~,\nn
\bar \gamma&=\frac{R_n^4}{64\N_c} \Bigl( \M_0^{(4)}(s_n)+\frac{4}{3} \M_2^{(4)}(s_n)\Bigr)~,
\end{align}
where $\M_I^{(n)}(s_n)$ denotes the $n$-th derivative of the function $\M_I(s)$, 
evaluated at the center of the Dalitz plot. 
Note that $\bar d$ and $\bar g$ do not receive contributions from the isospin $I=0$ amplitude.
\end{sloppypar}

Despite working in the limit of evaluating all amplitudes for the \emph{charged} pion mass, 
we employ the physical values for $R_c$ and $R_n$ in Eqs.~\eqref{eq:chargeddef}, \eqref{eq:neutraldef}.
These prefactors stem from the conversion of $\tilde a$, $\tilde \alpha$ etc.\ into
$\bar a$, $\bar \alpha$ etc., see Eq.~\eqref{eq:paramrel}, and are just due to a normalization choice
in the definition of the Dalitz plot variables $x$ and $y$; we therefore decide to present 
our results including this ``trivial'' isospin-breaking effect already at this stage.  
Note that due to $(\meta-3\mpn)/(\meta-3\mpc) \approx 1.11$ and
$(\meta-2\mpc-\mpn)/(\meta-3\mpc) \approx 1.04$, 
the effects of using these normalization factors in the isospin limit are large, 
most so for the neutral channel, where $\alpha$ for instance is affected by a shift of 22\%. 

In our numerical analysis we will observe that among the two-loop contributions 
those of the non-trivial two-loop graphs, see Fig.~\ref{fig:twolooptopologies} (right), 
are in general strongly suppressed. 
This can be traced back to the isospin properties of these pieces:
for those Dalitz plot parameters to which the $I=0$ partial wave can contribute, 
it usually dominates.  
For those graphs that only describe rescattering in one channel and can be written
as simple products of one-loop functions, see Fig.~\ref{fig:twolooptopologies} (middle),
the $I=0$ isospin amplitude receives contributions proportional to second powers of
$a_0$, $b_0$, etc., whereas the ``inner'' vertex in the non-trivial two-loop contributions
has parts of $I=0$ and $I=2$ (P-waves vanish due to symmetry reasons in the isospin limit)
that tend to partially cancel each other.  
\FIGURE[t]{
\centering
\includegraphics[width=0.95\linewidth]{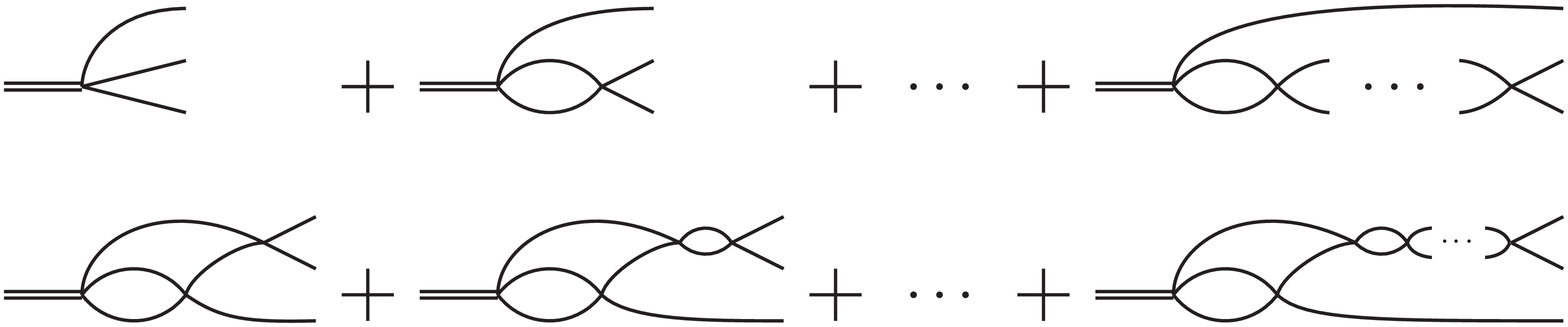}
\caption{Diagrammatic expression of the resummed amplitudes. Above: the bubble chain. Below: resummed external vertex of the non-trivial two-loop graph.
The line style is as in Fig.~\ref{fig:twolooptopologies}.}
\label{fig:iteratedbubbles}
}
In an attempt to estimate (partial) higher-order corrections, we
therefore expect to find a good approximation to the full result by iterating the bubble diagrams 
and the exterior two-particle rescattering of the non-trivial two-loop function 
as depicted diagrammatically in Fig.~\ref{fig:iteratedbubbles}. 
In the aforementioned representation the unitarized amplitudes are easily determined to be
\begin{align}
 \M_0^{\rm u}(s)&=\frac{\M_0(s)-\frac{5}{3}\ell_0(s)\bigl(16\pi a_0(s)J(s)\bigr)^2}{1-16\pi a_0(s)J(s)}~,\nn
 \M_1^{\rm u}(s)&=\frac{\M_1(s)}{1-16\pi a_1(s)q^2J(s)}~,\nn
 \M_2^{\rm u}(s)&=\frac{\M_2(s)-\ell_2(s)\bigl(16\pi a_2(s)J(s)\bigr)^2}{1-16\pi a_2(s)J(s)}~. \label{eq:unitarized}
\end{align}
The inclusion of iterated diagrams requires modified matching relations for the effective range parameters. This becomes obvious when 
considering the expansion of the iterated bubble sum of $\pi\pi$ scattering of isospin $I=0,2$ at the $\pi\pi$ threshold:
\begin{align}\label{eq:iterexp}
 {\rm Re}\Bigl[\frac{a_I(s)}{1-16\pi  a_I(s)J(s)}\Bigr]=a_I + \Bigl(b_I-\frac{a_I^3}{\mpc^2}\Bigr)q^2+\Bigl(c_I+\frac{a_I^3+a_I^5-3a_I^2b_I \mpc^2}{\mpc^4}\Bigr)q^4+\order(q^6)~.
\end{align}
One immediately sees that the effective range picks up a contribution from two-loop diagrams, the shape parameter from two- and four-loop diagrams. To account for this shift,
the above expression has to be compared with the effective range expansion of the $\pi\pi$ amplitude (for $l=0$),
\begin{align}\label{eq:effrexpamp}
 {\rm Re}\, t_0^I(q^2) =  a_I + b_I q^2 +c_Iq^4+\order(q^6)~,\qquad I=0,2~,
\end{align}
from which one reads off the following renormalization prescriptions:
\be
  a_I^{\rm ren}=a_I~,\qquad
  b_I^{\rm ren}=b_I+\frac{a_I^3}{\mpc^2}~,\qquad
  c_I^{\rm ren}=c_I-\frac{a_I^3-2 a_I^5-3 a_I^2b_I \mpc^2}{\mpc^4}~.
\ee
The $a_I^{\rm ren}$, $b_I^{\rm ren}$, $c_I^{\rm ren}$ are now to be inserted into the matching relations for the coupling constants 
$C_i$, $D_i$, $F_i$. The renormalization prescriptions have pretty remarkable 
effects in the isospin $I=0$ channel, where the shape parameter is shifted from $-0.002\mpc^{-4}$ to $+0.030\mpc^{-4}$ (for the ACGL parameter set). 
We note that the P-wave effective range $b_1$ does not pick up an additional
contribution due to the $q^{2}(s)$ prefactor. Corrections in the P-wave channel start at $\order(\epsilon^8)$, that is the higher-order shape parameter $d_1$.

\subsection{Numerical results}
\label{sec:numres}

\TABLE[t]{
\centering
\renewcommand{\arraystretch}{1.2}
\begin{tabular}{c r r r r r r r}
\toprule
			&\multicolumn{6}{l}{charged channel}				     \\
\midrule
			& $|\N_c|^2$	& $a\quad$ & $b\quad$& $d\quad$& $f\quad$ & $g\quad$ \\
\midrule
 \text{tree} 		& $0.0310$ 	& $-1.306$ & $0.393$ & $0.071$ & $0.022$  & $-0.046$ \\
 \text{one-loop} 	& $0.0338$ 	& $-1.450$ & $0.580$ & $0.085$ & $-0.026$ & $-0.078$ \\
 \text{two-loop*}	& $0.0289$ 	& $-1.288$ & $0.334$ & $0.093$ & $0.078$  & $-0.076$ \\
 \text{full two-loop} 	& $0.0287$ 	& $-1.290$ & $0.379$ & $0.056$ & $0.071$  & $-0.045$ \\
 \text{unitarized} 	& $0.0284$ 	& $-1.268$ & $0.342$ & $0.053$ & $0.101$  & $-0.042$ \\
\bottomrule
			&\multicolumn{3}{l}{neutral channel}\\
\midrule
			& $|\N_n|^2$	& $\alpha\quad$ & $\beta\quad$ 	& $\gamma\quad$ \\
\midrule
 \text{tree} 		& $0.279 $ 	& $0.0107$  	& $0\quad$	& $0.0001$ \\
 \text{one-loop} 	& $0.304 $ 	& $0.0227$  	& $0.0005$	& $0.0000$\\
 \text{two-loop*}	& $0.260 $ 	& $-0.0209$  	& $-0.0027$ 	& $0.0007$ \\
 \text{full two-loop} 	& $0.258 $ 	& $-0.0192$ 	& $-0.0036$	& $0.0009$ \\
 \text{unitarized} 	& $0.255 $ 	& $-0.0249$ 	& $-0.0043$	& $0.0013$ \\
\bottomrule
\end{tabular}
\renewcommand{\arraystretch}{1.0}
\caption{Results for the charged and neutral Dalitz plot parameters in the isospin limit. 
We show tree, one-loop, two-loop neglecting the irreducible two-loop graphs (marked ``two-loop*''), 
and full two-loop results calculated as described in the text, plus the result employing the unitarized amplitudes according to Eq.~\eqref{eq:unitarized}. 
\label{tab:DalitzT12u}}}

We begin our numerical analysis of the various $\eta\to3\pi$ Dalitz plot parameters by 
investigating how the tree-level values are modified at one- and two-loop order, and finally beyond
two loops (via the estimate through the unitarized amplitudes in Eq.~\eqref{eq:unitarized}).
This part of the analysis is based solely on the ACGL parameters for the $\pi\pi$ final-state interaction;
the qualitative conclusions are identical for the KPY parameterization.
We keep the $\eta\to3\pi$ tree level parameters fixed as obtained by matching to ChPT at $\Order(p^4)$
throughout, see Sect.~\ref{subsec:MatchingEta}.
Our results are summarized in Table~\ref{tab:DalitzT12u}.
In particular, we observe the following:
\begin{enumerate}
\item
Individual loop corrections to the Dalitz plot parameters are sizeable; their relative importance grows
with increasing order (in $\epsilon$) of the parameters concerned, as suggested a priori by 
power-counting arguments (see Sect.~\ref{subsec:powcounteta3pi}).
\item
One- and two-loop contributions are in general of the same size, as indeed expected, with a tendency
to cancel to varying extent due to contributions of opposite sign.  
This again substantiates the power-counting arguments of the NREFT framework,
which is particularly interesting in the case of $\alpha$: 
while at one loop we see a sizeable positive shift added to the already positive 
tree-level result, the two-loop correction overwhelms both, leading to a negative total.
We therefore find the correct sign for $\alpha$, as opposed to the ChPT result. 
At two loops our result is in fairly good agreement with the dispersive one from Ref.~\cite{KWWeta3pi}. 
\item
There are large contributions from derivative couplings at two-loop order. This is seen when considering the amplitude 
expanded only up to $\order(a_{\pi\pi}\epsilon^5,a_{\pi\pi}^2\epsilon^2)$ (cf.\ Ref.~\cite{KL3pi}), 
at which order only constant vertices are implemented at two loops. 
In this approximation, we find numerically e.g.\ $\alpha=+0.033$. Once the effective range corrections in the $I=0$ 
two-loop bubble are added, $\alpha$ receives a shift to $-0.017$. 
This observation explains why the authors of Ref.~\cite{Gullstrom} obtain a positive sign for $\alpha$ when matching to
ChPT at tree-level: no derivative couplings at two-loop level are included in that work. With respect to this omission,
matching to ChPT at tree-level plays a minor role in the deviation from our result.
\item
By comparing the two-loop contributions with and without the parts due to the irreducible two-loop graphs,
see Fig.~\ref{fig:twolooptopologies} (right), we see that at least in those parameters that receive
contributions from the $I=0$ amplitude the irreducible two-loop graphs only give a very
small contribution.  As detailed before, this can be traced back to the isospin structure of the
different amplitudes.  Specifically $\alpha$ is a case in point: the simple ``bubble sum'' type
two-loop graphs shift it by about $-0.044$, while the irreducible graphs only add $+0.002$.
\item
Our estimate of higher-order effects via simple two-channel unitarization shows that those are 
significantly smaller than the (individual) one- and two-loop effects, although not negligible throughout.
Due to the smallness of the irreducible two-loop graphs, we expect to catch the major part
of the higher-order corrections in this way.
\end{enumerate}

\TABLE[t]{
\centering
\renewcommand{\arraystretch}{1.2}
\begin{tabular}{c r r r r rcl}
\toprule
	    & \text{ACGL 2-loop}& \text{ACGL unit.}& \text{KPY 2-loop}	& \text{KPY unit.} & 		&    &\hspace{-1.2cm}\text{average} \\
\midrule
$|\N_c|^2$&$ 0.0287\ \quad$	&$ 0.0284\ \quad$	&$ 0.0285\quad$		&$ 0.0282 \quad$	&$ 0.0284$	&\tpm&\tbk$ 0.0002$ \\
 $a$      &$-1.290\ \quad$ 	&$-1.268\ \quad$	&$-1.291\quad$ 		&$-1.267 \quad$ 	&$-1.279$	&\tpm&\tbk$ 0.012 $ \\
 $b$      &$ 0.379\ \quad$ 	&$ 0.342\ \quad$	&$ 0.382\quad$ 		&$ 0.340 \quad$ 	&$ 0.361 $	&\tpm&\tbk$ 0.021 $ \\
 $d$      &$ 0.056\ \quad$ 	&$ 0.053\ \quad$	&$ 0.052\quad$ 		&$ 0.050 \quad$ 	&$ 0.053 $	&\tpm&\tbk$ 0.003 $ \\
 $f$      &$ 0.071\ \quad$ 	&$ 0.101\ \quad$	&$ 0.073\quad$ 		&$ 0.107 \quad$ 	&$ 0.089$	&\tpm&\tbk$ 0.018 $ \\
 $g$      &$-0.045\ \quad$ 	&$-0.042\ \quad$	&$-0.043\quad$ 		&$-0.041 \quad$ 	&$-0.043$	&\tpm&\tbk$ 0.002 $ \\
 $\alpha$ &$-0.0192\ \quad$	&$-0.0249\ \quad$	&$-0.0227\quad$		&$-0.0291 \quad$	&$-0.0242$	&\tpm&\tbk$ 0.0049$ \\
 $\beta$  &$-0.0036\ \quad$	&$-0.0043\ \quad$	&$-0.0043\quad$		&$-0.0051 \quad$	&$-0.0043$	&\tpm&\tbk$ 0.0007$ \\
 $\gamma$ &$ 0.0009\ \quad$	&$ 0.0013\ \quad$	&$ 0.0013\quad$		&$ 0.0017 \quad$	&$ 0.0013$	&\tpm&\tbk$ 0.0004$ \\
\bottomrule
\end{tabular}
\renewcommand{\arraystretch}{1.0}
\caption{Results for charged and neutral Dalitz plot parameters in the isospin limit with different
input on $\pi\pi$ scattering parameters from Refs.~\cite{ACGL} and \cite{Yndurain}; 
see Sect.~\ref{subsec:MatchingPipi}.  Shown are the results both for two loops and for the unitarized amplitudes. 
$|\N_n|^2=9|\N_c|^2$ is not shown separately. \label{tab:DalitzPipi}}
}

In order to study the dependence of our results on the precise input for $\pi\pi$ scattering,
we next compare the values obtained for the various charged and neutral Dalitz plot parameters,
at two loops and unitarized, for the ACGL and the KPY parameter sets in Table~\ref{tab:DalitzPipi}. 
In most cases, the variation with different $\pi\pi$ input is a bit smaller than the difference
due to the higher-order estimates, although not by much.
As our final result in the last column of Table~\ref{tab:DalitzPipi}, we determine
central values and (symmetric) errors in such a way as to cover all four values for each parameter.

Our finding for the $\eta\to 3\pi^0$ slope parameter, $\alpha = -0.024 \pm 0.005$, 
is considerably closer to the current experimental average $\alpha = -0.0317 \pm 0.0016$~\cite{pdg} than previous theoretical approaches.  
Note again that the theoretical prediction for $\alpha$ is lowered (in absolute value) by about 22\% if 
the charged pion mass is used in the definition of $z$.
We predict the (yet unmeasured) higher-order Dalitz plot parameters $\beta$ and $\gamma$ in the neutral channel to 
be different from zero, but very small.  
In particular, neglecting a term $\propto \gamma z^2$ in an experimental extraction of $\alpha$
based on the radial distribution $d\Gamma/dz$ alone (in which a term $\propto \beta$ cancels
for $z<0.756$, compare Fig.~\ref{fig:cusps})
should affect $\alpha$ by less than the value of $\gamma$, hence still below the current uncertainty,
although not by much given the precision of the most recent experimental determinations.

As we will see below, there are sizeable isospin-breaking shifts in the charged Dalitz plot parameters.
We therefore defer a detailed comparison to experimental values to Sect.~\ref{sec:isospinbreaking}.
We only wish to make a remark on the cubic parameters $f$ and $g$ here.
Apart from the fact that a large contribution to these is given by $2{\rm Re}(\bar a \bar b^*)$ and 
$2{\rm Re}(\bar a \bar d^*)$, respectively, the remainders (or $\bar f$, $\bar g$) are given
entirely in terms of loop contributions.  
If we, in addition, allow for cubic tree level terms $\bar f^\tr$ and $\bar g^\tr$ 
and match the latter to ChPT at $\Order(p^4)$, the total results receive 
shifts of $-0.002$ and $-0.011$ hence very and relatively small effects, respectively.
Although chiral $\Order(p^6)$ corrections might modify these numbers significantly, 
we still regard them as indications that the dominance of loop contributions
(as suggested by $\epsilon$ power counting) holds here.

\subsection[Comparison to $\alpha$ in ChPT at two loops]{Comparison to \boldmath{$\alpha$} in ChPT at two loops}

While dispersive analyses find values for $\alpha$ similar to ours~\cite{KWWeta3pi},
a serious puzzle is the question why the calculation of this quantity
in ChPT to two loops~\cite{BGeta3pi} does not arrive at least at a negative value for $\alpha$
``naturally'', i.e.\ as the central value (disregarding the large error bar due to the estimated
fit uncertainty).  
After all, in addition to potentially significant chiral SU(3) renormalization effects 
of what would be subsumed in the tree-level couplings of the NREFT representation, ChPT at $\Order(p^6)$ also
includes all the pion two-loop graphs shown to be important here.

It turns out that this failure of the chiral two-loop calculation can partly be understood within our framework,
investigating rescattering effects only, but of course neglecting the $\order(p^6)$ modified tree-level couplings.
In order to mimic the chiral expansion,
we note that in an $\Order(p^6)$ calculation, 
the $\pi\pi$ vertices inside two-loop graphs are only included to their current-algebra (or $\Order(p^2)$) accuracy,
see Eq.~\eqref{eq:CApipi}, while inside the one-loop diagrams, $\pi\pi$ rescattering is taken care of
up to $\Order(p^4)$.  By inserting the respective values for the $\pi\pi$ threshold parameters
in our amplitude, we find 
\be \label{eq:alphaSimOp6}
 \alpha_{\text{ChPT}} = - 0.0011~,
\ee
hence a value close to zero.  We attribute the remaining difference to the central result for 
$\alpha$ in Ref.~\cite{BGeta3pi} to different tree-level couplings as determined in that paper.
As we found that precisely the two-loop effects turn $\alpha$ negative, dominated by the $I=0$ amplitude,
we conclude that a large part of the discrepancy between ChPT at $\Order(p^6)$ and our result
(or the one from dispersion relations) is due to the significantly weaker $\pi\pi$ rescattering
(compare e.g.\ $a_0^\CA\approx 0.16$ vs.\ $a_0=0.220$ from Ref.~\cite{ACGL}, which enters the two-loop effects
squared).
The precise choice of the set of rescattering parameters therefore has a large effect on the result for $\alpha$ 
(and, slightly less dramatically so, on other Dalitz plot parameters). 
The inclusion of improved values for the effective ranges and shape parameters produces a large shift of the 
chiral result towards the experimental value. 

In a very condensed manner, we can therefore point to one specific diagram, Fig.~\ref{fig:twolooptopologies} (middle),
which accounts for roughly half of the discrepancy between the central value of the chiral prediction at 
$\Order(p^6)$ and the experimental value for $\alpha$.
More specifically, the discrepancy is caused by contributions of the diagrammatic topology of this kind. 
Since (at least) next-to-leading order contributions to the $\pi\pi$ vertices are required, 
one needs to include these diagrams up to $\order(p^8)$ and higher in strict chiral power counting.
To substantiate this claim and ensure that it is not an artifact of the non-relativistic framework,
we replace the non-relativistic two-point function $J(s)$, Eq.~\eqref{eq:loopJ}, by its relativistic
counterpart $\bar J_{\pi\pi}(s)$, which differs from the former by its real part
(given explicitly in Eq.~\eqref{eq:Jbar}).
Doing so requires a different matching procedure to account for the (otherwise absent) mass
renormalization effects on the various coupling constants thus induced; we will not spell out 
this exercise in detail.
The main conclusion however is fully consistent with our findings above:
the ``double bubble'' graphs alone shift $\alpha$ by $-0.042$ (to be compared with $-0.044$, see Table~\ref{tab:DalitzT12u});
calculating them with current algebra values for the $\pi\pi$ threshold parameters reduces this 
effect by nearly a factor of two, which corresponds to the discrepancy between Eq.~\eqref{eq:alphaSimOp6}
and the value obtained in NREFT.

One might argue that a parameter as subtle as $\alpha$ could also be subject to other very sizeable $\order(p^6)$ corrections; 
in particular, contributions from chiral low-energy constants appear for the first time at that order.
For a superficial impression of these effects, we investigate precisely the 
$\order(p^6)$ polynomial in the amplitude calculated in Ref.~\cite{BGeta3pi}. 
One easily finds the following combination of low-energy constants contributing to $\alpha$:
\begin{align}\label{eq:p6LECs}
 \alpha_{\rm LEC}^{(6)}=\frac{12R_n^2}{F_{\pi}^4}\big(C_5^r + C_8^r + 3 C_9^r+C_{10}^r - 2 C_{12}^r + 2 C_{22}^r + 3 C_{24}^r + C_{25}^r\big)~.
\end{align}
The couplings $C_i^r$  are estimated in Ref.~\cite{BGeta3pi} using resonance saturation.
Vector contributions cancel in Eq.~\eqref{eq:p6LECs}, as they must, with no P-waves appearing in the neutral
decay channel. Using the scalar resonance estimates given in Ref.~\cite{BGeta3pi}, we arrive at
the very simple and compact expression
\begin{align}\label{eq:p6ResSat}
 \alpha_{\rm LEC}^{(6)}=\frac{12 R_n^2 c_d c_m}{F_{\pi }^2 M_S^4} \approx 0.005~,
\end{align}
where $c_m=0.042$~GeV, $c_d=0.032$~GeV, and $M_S=0.98$~GeV. 
There are serious doubts about the reliability of the resonance saturation hypothesis in the 
scalar sector~\cite{Leutwyler:2008ma}; indeed on might argue that the masses of even heavier
scalar states ought to be used in Eq.~\eqref{eq:p6ResSat}, further suppressing 
their contribution to $\alpha$.
We nevertheless confirm that contributions from the ChPT low-energy polynomial at $\order(p^6)$ are rather small;
in particular they have a positive sign, so they cannot serve as an alternative explanation to arrive
at a negative $\alpha$. 
We also emphasize that the above is only a very rough estimate of the expected size of the effects 
and does not by any means replace a consistent matching procedure.

\section{Isospin breaking in \boldmath{$\eta\to 3\pi$}}
\label{sec:isospinbreaking}
In this section we discuss higher-order isospin-breaking contributions to the decay $\eta\to 3\pi$. 
We concentrate on the following four contributions:
\begin{enumerate}
\item
Isospin breaking in $\eta\to\pi^+\pi^-\pi^0$ due to $Q_n \neq Q_c$. 
There are significant corrections to the charged Dalitz plot parameters due to the terms $\propto (R_n-R_c)$ in Eq.~\eqref{eq:paramrel},
which stem from the subtleties in the definition of the center of the Dalitz plot discussed in Sect.~\ref{sec:Dalexp}.
\item 
Other isospin corrections due to the difference between the charged and the neutral pion mass. 
These in particular concern the incorporation of the correct thresholds inside the loop contributions,
which is necessary for a description of the boundary regions of the Dalitz plot, among them
the cusp effect in $\eta\to3\pi^0$.
The representation of the amplitude in the non-relativistic framework allows us to work in the particle 
(and not in the isospin) basis, and thus we can incorporate mass effects in a straightforward fashion.
\item 
Isospin-breaking corrections to the $\pi\pi$ rescattering parameters. 
We use the phenomenological values for the scattering lengths and effective ranges, 
which have been determined in the isospin limit~\cite{ACGL,Yndurain},
and calculate corrections to each channel from the one-loop SU(2) $\pi\pi$-scattering amplitudes 
with electromagnetic corrections included.
\item \label{item:isotree}
Next-to-leading-order isospin-breaking effects in the $\eta\to3\pi$ tree level couplings,
calculated in one-loop ChPT,
which modify Eqs.~\eqref{eq:Dalrel}.
\end{enumerate}
The representation of the NREFT amplitude to two loops with fully general masses and coupling constants,
allowing for all of these isospin-breaking effects, is given in Appendix~\ref{app:NREFT2loop}.
Furthermore Appendix~\ref{app:NREFTunit} shows the generalization of the unitarization prescription given in
Eq.~\eqref{eq:unitarized} for the case of isospin violation.
In our numerical evaluation we will add these contributions cumulatively to the results of Sect.~\ref{sec:isospinlimitresults}.

In this context we should comment on radiative (real- and virtual-photon) corrections to these decays.
In order to be able to sensibly discuss a Dalitz plot expansion of the squared amplitudes in question, 
we assume that the universal
radiative corrections (Gamow--Sommerfeld factor, bremsstrahlung contributions etc.),
as discussed in the framework of NREFT in Ref.~\cite{RadK3pi}, have already been subtracted 
from the experimental data when determining Dalitz plot parameters.  
In order to extract the corrections of point~\ref{item:isotree} above from the calculation
in Ref.~\cite{DKMeta3pi}, these subtracted contributions have to be matched correctly,
as detailed in Appendix~\ref{app:isosprelcorr}.
The non-universal or ``internal'' radiative corrections that play an important role
in the analysis of the cusp effect in $K\to3\pi$~\cite{RadK3pi,Batley:2000zz} 
do not have a similarly enhanced effect in the center of the Dalitz plot.
From the point of view of chiral power counting of isospin-breaking corrections, 
these constitute higher-order effects than those considered consistently in Ref.~\cite{DKMeta3pi}
(as they only appear at two loops);
furthermore, in $\eta\to\pi^+\pi^-\pi^0$, even diagrams beyond those calculated in Ref.~\cite{RadK3pi}
would have to be included.  
We have checked, though, that the effect of photon exchange inside the charged-pion loops
on the $\eta\to3\pi^0$ Dalitz plot expansion is small, even on the scale of the other small isospin-breaking 
effects discussed below.

The by far largest isospin-breaking effects on the Dalitz plot parameters,
beyond the use of the correct overall normalization factors of $Q_n$ and $Q_c$ in the definitions
of the kinematic variables that was already incorporated in the previous sections, 
are the kinematic effects due to the fact that for the decay $\eta\to\pi^+\pi^-\pi^0$
the position defined by $x=y=0$ does not coincide with $s_1=s_2=s_3$ when $\mpc\neq\mpn$. 
Using the correct prescriptions given in Eq.~\eqref{eq:paramrel}, we find the results listed in the left column of Table~\ref{tab:isobreak}. The corrections are very sizeable: our analysis shows that $a$ is reduced 
(in magnitude) by 5\%, $b$ by even 14\%. These kinematic effects constitute the bulk of the isospin breaking corrections to the charged parameters.

\TABLE[t]{
\centering
\renewcommand{\arraystretch}{1.2}
\begin{tabular}{c rcl rcl rcl rcl}
\toprule
	  & $Q_c$	&\tbk$\neq$&\tbk$Q_n$	&	&    &\hspace{-1.2cm}masses 	& 		&    &\hspace{-0.85cm}$\pi\pi$ 	& 		&    &\hspace{-1.2cm}$\eta\to 3\pi^0$ \\
\midrule
$|\N_c|^2$& $ 0.0310$	&\tpm&\tbk$0.0003$ 		& $ 0.0309$	&\tpm&\tbk$0.0003$ 		& $ 0.0310$	&\tpm&\tbk$0.0003$ 		& $ 0.0310$	&\tpm&\tbk$0.0003$ \\
$a $      & $-1.218$	&\tpm&\tbk$0.013 $ 		& $-1.214$	&\tpm&\tbk$0.013 $ 		& $-1.214$	&\tpm&\tbk$0.014 $ 		& $-1.213$	&\tpm&\tbk$0.014 $ \\
$b $      & $ 0.314$	&\tpm&\tbk$0.023 $ 		& $ 0.310$	&\tpm&\tbk$0.023 $ 		& $ 0.308$	&\tpm&\tbk$0.023 $ 		& $ 0.308$	&\tpm&\tbk$0.023 $ \\
$d $      & $ 0.051$	&\tpm&\tbk$0.003 $ 		& $ 0.051$	&\tpm&\tbk$0.003 $ 		& $ 0.050$	&\tpm&\tbk$0.003 $ 		& $ 0.050$	&\tpm&\tbk$0.003 $ \\
$f $      & $ 0.084$	&\tpm&\tbk$0.019 $ 		& $ 0.082$	&\tpm&\tbk$0.018 $ 		& $ 0.083$	&\tpm&\tbk$0.019 $ 		& $ 0.083$	&\tpm&\tbk$0.019 $ \\
$g $      & $-0.039$	&\tpm&\tbk$0.002 $ 		& $-0.039$	&\tpm&\tbk$0.002 $ 		& $-0.039$	&\tpm&\tbk$0.002 $ 		& $-0.039$	&\tpm&\tbk$0.002 $ \\
$|\N_n|^2$& $ 0.256$	&\tpm&\tbk$0.002 $ 		& $ 0.256$	&\tpm&\tbk$0.002 $ 		& $ 0.255$	&\tpm&\tbk$0.002 $ 		& $ 0.252$	&\tpm&\tbk$0.008 $ \\
$\alpha$  & $-0.0242$	&\tpm&\tbk$0.0049$ 		& $-0.0241$	&\tpm&\tbk$0.0049$ 		& $-0.0247$	&\tpm&\tbk$0.0048$ 		& $-0.0246$	&\tpm&\tbk$0.0049$ \\
$\beta$	  & $-0.0043$	&\tpm&\tbk$0.0007$ 		& $-0.0043$	&\tpm&\tbk$0.0008$ 		& $-0.0042$	&\tpm&\tbk$0.0007$ 		& $-0.0042$	&\tpm&\tbk$0.0007$ \\
$\gamma$  & $ 0.0013$	&\tpm&\tbk$0.0004$ 		& $ 0.0012$	&\tpm&\tbk$0.0004$ 		& $ 0.0013$	&\tpm&\tbk$0.0004$ 		& $ 0.0013$	&\tpm&\tbk$0.0004$ \\
\bottomrule
\end{tabular}
\renewcommand{\arraystretch}{1.0}
\caption{Central results for the charged and neutral Dalitz plot parameters with isospin breaking in 
kinematic relations, masses in loop functions, 
$\pi\pi$ threshold parameters, and $\eta\to 3\pi^0$ tree level couplings (see text for details).
\label{tab:isobreak}}
}

The modifications that arise from using physical pion masses 
in the loop functions and derivative couplings are very small in the expansion around the center of the Dalitz plot, 
see Table~\ref{tab:isobreak} (second column).  
The charged parameters are typically reduced in magnitude on the level of about 1\%;
$\alpha$ is shifted by $+ 0.0001$ only, an order of magnitude below 
the uncertainty due to different $\pi\pi$ parametrizations.  
The importance of pion-mass effects in loops only becomes really visible when studying the full Dalitz plot 
distribution also at its boundaries (see Refs.~\cite{DKMeta3pi,Gullstrom}).

The next column in Table~\ref{tab:isobreak} shows the effect of isospin-breaking corrections
in the $\pi\pi$ threshold parameters.  For this purpose, 
we have calculated the electromagnetic contributions to the matching relations
up-to-and-including $\Order(e^2p^2)$ in the chiral expansion
for S- and P-wave scattering lengths and S-wave effective ranges,
using the results for the one-loop $\pi\pi$ scattering amplitudes in the presence
of virtual photons of Refs.~\cite{KnechtUrech,KnechtNehme}.
The necessary matching procedure is described in detail in Appendix~\ref{app:EMcorr}.
The modifications in the Dalitz plot parameters
are largest for $\alpha$, where a $3\%$ effect is observed. 
The contributions to the remaining parameters stay well below or around $1\%$.
In all cases the shifts are dominated by isospin-breaking corrections in the S-wave scattering
lengths and thereby the $\Order(e^2)$ chiral corrections, as expected by power counting.

Finally, we want to investigate the effects of isospin breaking on the relations in Eq.~\eqref{eq:Dalrel},
i.e.\ next-to-leading order isospin breaking in the $\eta\to 3\pi$ tree level couplings.
These can be extracted from the chiral one-loop calculation of the $\eta\to3\pi$ decay amplitudes
to $\Order(e^2(m_d-m_u))$ in Ref.~\cite{DKMeta3pi}.
We write the corrections in the form
\be\label{eq:Dalrelcorr}
  \N_n = -3\tilde \N_c + \Delta_{\tilde \N} ~, \qquad 
  \tilde\alpha =\frac{1}{3}(\tilde b+3\tilde d) + \Delta_{\tilde \alpha}~,
\ee
where $\Delta_{\tilde \N} = \Order(e^2(m_d-m_u))$ and $\Delta_{\tilde \alpha} = \Order(e^2)$.
Note that no corrections of $\Order((m_d-m_u)^2)$ (in $\Delta_{\tilde \N}$) 
and $\Order(m_d-m_u)$ (in $\Delta_{\tilde \alpha}$) occur, respectively.
The analytic results of the expansion and further details are given in Appendix~\ref{app:isosprelcorr}. 
With the numerical input for various low-energy constants chosen as in Ref.~\cite{DKMeta3pi},
we find that the corrections to the isospin relations are very small,
\begin{align}\label{eq:Dalrelcorrnumres}
  \frac{\Delta_{\tilde \N}}{\N_n}=(-0.7\pm1.5)\%~,\qquad \Delta_{\tilde\alpha}=0.035\pm0.003~{\rm GeV}^{-4}~.
\end{align}
The numerical analysis shows that the corrections to $\alpha$ are below $1\%$ and thus 
very small, even for isospin breaking corrections.
It is interesting to note that the modification induced by $\Delta_{\tilde\alpha}$ is largely counterbalanced by the 
modification due to $\Delta_{\tilde \N}$. 
Even though the modifications Eq.~\eqref{eq:Dalrelcorr} only affect the $\eta\to3\pi^0$
tree-level couplings, these in principle also enter the charged channel via (inelastic) rescattering effects, 
however these shifts are too small to register.
The corresponding values, which also constitute our final results, are collected in 
the final column of Table~\ref{tab:isobreak}.

\TABLE[t]{
\centering
\renewcommand{\arraystretch}{1.2}
\begin{tabular}{c rcl rcl rcl}
\toprule
Theory							&&\tbk$a$&							&&\tbk$b$&						&&\tbk$d$&					\\
\midrule
ChPT $\order(p^4)$					&$-1.34$&\tpm&\tbk$0.04$					&$0.434$&\tpm&\tbk$0.018$				&$~~0.077$&\tpm&\tbk$0.008$			\\
ChPT $\order(p^6)$					&$-1.271$&\tpm&\tbk$0.075$					&$0.394$&\tpm&\tbk$0.102$				&$~~0.055$&\tpm&\tbk$0.057$			\\
Dispersive						&$-1.16$&&							&$0.24$&\tbk\ldots&\tbk$0.26$				&$~~0.09$&\tbk\ldots&\tbk$0.10$			\\
$\order(p^4)$+NREFT					&$-1.213$&\tpm&\tbk$0.014$					&$0.308$&\tpm&\tbk$0.023$				&$~~0.050$&\tpm&\tbk$0.003$			\\
\toprule
Experiment						&&\tbk$a$&					&&\tbk$b$&				&&\tbk$d$						\\
\midrule
  KLOE~\cite{KLOEcharged}				&$-1.090$&\tpm&\tbk$0.005_{-0.019}^{+0.008}$\hspace{-1mm}	&$0.124$&\tpm&\tbk$0.006\pm0.010$\hspace{-4mm}		&$~~0.057$&\tpm&\tbk$0.006_{-0.016}^{+0.007}$	\\
Crystal Barrel~\cite{CBcharged}	\hspace{-4mm}		&$-1.22$&\tpm&\tbk$0.07$					&$0.22$&\tpm&\tbk$0.11$					&$~~0.06$&\tpm&\tbk$0.04$ (input)		\\
Layter {\it et al.}~\cite{Laytercharged}		&$-1.08$&\tpm&\tbk$0.014$					&$0.034$&\tpm&\tbk$0.027$				&$~~0.046$&\tpm&\tbk$0.031$			\\
Gormley {\it et al.}~\cite{Laytercharged}\hspace{-4mm}	&$-1.17$&\tpm&\tbk$0.02$					&$0.21$&\tpm&\tbk$0.03$					&$~~0.06$&\tpm&\tbk$0.04$			\\
\bottomrule
\end{tabular}
\renewcommand{\arraystretch}{1.0}
\caption{Results for the charged Dalitz plot parameters in comparison with various theoretical and experimental determinations (the next-to-leading-order errors are only due to $L_3^{\rm r}$).\label{tab:finalresultscharged}}
}
After analyzing the isospin-breaking contributions we can now compare our final results for the charged Dalitz plot parameters with several other theoretical determinations and experimental findings in Table~\ref{tab:finalresultscharged}. 

We receive mixed results for the different Dalitz plot parameters. While $d$ is in good agreement with experiment, $a$ shows deviations of about 10\% to the results from Layter and most notably from the 
precision measurement of the KLOE collaboration, which -- due to the relatively small errors -- exceeds even very generous confidence levels. Our result is more or less compatible with the $\order(p^6)$
ChPT result. The dispersive calculation is somewhat closer to experiment, but no error range is given for us to compare with. The situation is even worse with $b$, where the deviation between our 
result and the KLOE measurement is rather alarming. The dispersive analysis indicates that even higher-order effects might be somewhat important in the determination of $a$ and $b$, however it cannot account for the
discrepancy we find for $b$. A main source of uncertainty that we have not addressed so far is the tree-level input, which could receive rather large contributions from matching to the chiral amplitude at 
$\order(p^6)$. It is possible that the deviation in $a$ can be accounted for by such a matching prescription. There is no indication, however, that this is also the case for $b$.
This issue is put under tense scrutiny in the next section. The results obtained in that discussion question to some extent the consistency between the charged and neutral Dalitz plot measurements.
---
Our result for the cubic parameter, $f=0.083\pm0.019$, is reasonably compatible with the KLOE determination, $f=0.14\pm0.01\pm0.02$~\cite{KLOEcharged}.

\section{Relating charged and neutral Dalitz plot parameters}
\label{sec:RelatingDal}

\subsection[Isospin limit $Q_n=Q_c$]{Isospin limit \boldmath{$Q_n=Q_c$}}

\TABLE[t]{
\centering
\renewcommand{\arraystretch}{1.2}
\begin{tabular}{c rcl rcl rcl rcl rcl rcl}
\toprule
 \!\text{Input} \!& &\tbk 1&\tbk 			& 	   &\tbk$\bar a$&\hspace{-4.5mm}$^\tr$	&	  &\tbk$\bar b$&\hspace{-4.5mm}$^\tr$  	&	  &\tbk$\bar d$&\hspace{-4.5mm}$^\tr$	\\
 \midrule
 $\N/\tilde{\N}_c\!$ & $ 0.9119$&\tbk$+$&\tbk$0.2954 i$	& $ 0.0028$&\tbk$+$&\tbk$0.0005 i$ 		& $-0.0097$&\tbk$+$&\tbk$0.0174 i$  		& $-0.0156$&\tbk$+$&\tbk$0.0643 i$ \\
 $\bar a$    		& $ 0.0202$&\tbk$-$&\tbk$0.4228 i$	& $ 1.0092$&\tbk$-$&\tbk$0.1902 i$ 		& $-0.0393$&\tbk$-$&\tbk$0.0182 i$ 		& $-0.0200$&\tbk$-$&\tbk$0.0378 i$ \\
 $\bar b$    		& $-0.0421$&\tbk$-$&\tbk$0.0166 i$	& $ 0.0152$&\tbk$-$&\tbk$0.1205 i$ 		& $ 1.0106$&\tbk$-$&\tbk$0.0834 i$  		& $-0.0069$&\tbk$+$&\tbk$0.0079 i$ \\
 $\bar d$    		& $-0.0182$&\tbk$+$&\tbk$0.0127 i$	& $-0.0156$&\tbk$-$&\tbk$0.0483 i$ 		& $ 0.0091$&\tbk$-$&\tbk$0.0079 i$  		& $ 0.9782$&\tbk$-$&\tbk$0.3583 i$ \\
 $\bar f$    		& $-0.0009$&\tbk$-$&\tbk$0.0118 i$	& $-0.0327$&\tbk$+$&\tbk$0.0011 i$ 		& $ 0.0331$&\tbk$-$&\tbk$0.2371 i$  		& $-0.0214$&\tbk$-$&\tbk$0.1175 i$ \\
 $\bar g$    		& $ 0.0041$&\tbk$-$&\tbk$0.0027 i$	& $-0.0031$&\tbk$+$&\tbk$0.0074 i$ 		& $-0.0022$&\tbk$-$&\tbk$0.0115 i$ 		& $-0.0330$&\tbk$-$&\tbk$0.0783 i$ \\
 $\bar\alpha$		& $-0.0345$&\tbk$-$&\tbk$0.0028 i$	& $-0.0004$&\tbk$-$&\tbk$0.0964 i$ 		& $ 0.5823$&\tbk$-$&\tbk$0.0522 i$  		& $ 0.5548$&\tbk$-$&\tbk$0.2001 i$ \\
 $\bar\beta$ 		& $ 0.0015$&\tbk$+$&\tbk$0.0028 i$	& $ 0.0090$&\tbk$+$&\tbk$0.0018 i$ 		& $-0.0108$&\tbk$+$&\tbk$0.0688 i$  		& $-0.0036$&\tbk$+$&\tbk$0.0119 i$ \\
 $\bar\gamma$		& $-0.0010$&\tbk$-$&\tbk$0.0064 i$	& $-0.0008$&\tbk$-$&\tbk$0.0016 i$ 		& $-0.0216$&\tbk$-$&\tbk$0.0016 i$  		& $-0.0075$&\tbk$+$&\tbk$0.0099 i$ \\
\bottomrule
\end{tabular}
\renewcommand{\arraystretch}{1.0}
\caption{Parameterization of Dalitz plot parameters in terms of tree input parameters.\label{tab:param}}
}

Up to this point we have only discussed uncertainties due to the effective range parameterization in the final-state interactions. 
A by far greater source of uncertainty is the tree-level input to our calculation, i.e.\ the matching to the ChPT one-loop 
amplitude, which we deem responsible for most of the remaining deviation from the experimental results. 
At higher orders (chiral $\Order(p^6)$), these tree parameters will receive chiral SU(3) corrections,
or renormalizations of $\Order(m_s)$, which certainly are potentially large.
In order to document our findings beyond the matching to the chiral one-loop amplitude,
we provide a direct parameterization of the various Dalitz plot parameters in terms of these input (tree) parameters. 
For this purpose, we first revert back to the case $Q_n=Q_c$ as expressions become much simpler in this limit.
Table~\ref{tab:param} shows the coefficients of the respective input parameters.
The entries are to be understood as follows: e.g., the second line means that the value
for $\bar a$ including final-state interactions is determined by the tree input according to
\begin{align}
\bar a &= 0.0202 -0.4228 i + (1.0092 -0.1902 i)\bar a^\tr  \nn
&\quad - (0.0393+0.0182 i)\bar b^\tr - (0.0200+0.0378 i)\bar d^\tr~.
\label{eq:aDep}
\end{align}
All numerical coefficients are determined by $\pi\pi$ scattering alone.
They are again averaged from the four different results (two loops plus unitarized, with 
ACGL and KPY parameters used as input) as in Table~\ref{tab:DalitzPipi}.
The error range of Dalitz plot parameters calculated with this parameterization may be taken from the 
last column of Table~\ref{tab:DalitzPipi}.
Furthermore, we only show the relations linear in the tree parameters
(that is, no terms of quadratic etc.\ order), which are the by far dominant contributions.

As we will now show, Table~\ref{tab:param} can be used to construct an explicit relation between
charged and neutral channel Dalitz plot parameters.
From Eq.~\eqref{eq:Dalrel} one can derive the following relation (again, we only consider
$R_n \neq R_c$ or $Q_n \neq Q_c$ in the overall normalization for the moment):
\be \label{eq:alphaisosp}
  \alpha=\frac{Q_n^2}{4Q_c^2}\big(d+b- |\bar a|^2 \big)~,
\ee
and consequently (cf.\ Ref.~\cite{BGeta3pi})
\be \label{eq:alphaineq}
  \alpha = \frac{Q_n^2}{4Q_c^2}\Big(d+b-\frac{a^2}{4} \Big) - \frac{Q_n^2}{4Q_c^2} \big({\rm Im}(\bar a)\big)^2
\leq \frac{Q_n^2}{4Q_c^2}\Big(d+b-\frac{a^2}{4} \Big)~,
\ee
which turns into an equality only for $\rm{Im}(\bar a)=0$.
The obvious question arises: as $\rm{Im}(\bar a)$ is generated by final-state interactions 
but in turn depends on the Dalitz plot parameters, can we quantify the \emph{equality} in Eq.~\eqref{eq:alphaineq}
in such a way that we obtain a testable consistency relation between the experimental observables
$\alpha$, $a$, $b$, and $d$, \emph{independent} of any (potentially insufficiently accurate) ChPT input?
The answer is yes -- precisely by using the information contained in Table~\ref{tab:param}.
We consider Eq.~\eqref{eq:aDep} and first note that, to very good accuracy, 
the contributions from $\bar b^\tr$ and $\bar d^\tr$ can be neglected:
with these parameters matched as previously, we have 
$\bar a^\tr\approx -0.656$,
$\bar b^\tr\approx -0.017$,
$\bar d^\tr\approx  0.037$,
which is sufficient to demonstrate that $\bar b^\tr$ and $\bar d^\tr$ are suppressed
compared to $\bar a^\tr$ by at least one order of magnitude, irrespective of potential
higher-order corrections.
(The neglected terms are retained explicitly in the following Sect.~\ref{sec:RelatingDalIsobreak}, 
compare Eq.~\eqref{eq:alphashifted}, which fully justifies their omission.)
So via $\bar a^\tr$ in Eq.~\eqref{eq:aDep}, ${\rm Im}(\bar a)$ can be solved for 
${\rm Re}(\bar a) = a/2$, and we find 
\be \label{eq:alphaisospnum}
\alpha=\frac{Q_n^2}{4Q_c^2}\Bigl(b+d-\frac{a^2}{4}\Bigr) - \zeta_1(1+\zeta_2 a)^2 ~, \quad
\zeta_1=0.050\pm 0.005 ~, \quad
\zeta_2=0.225\pm 0.003 ~.
\ee
We wish to emphasize once more that the values for $\zeta_{1/2}$ depend solely on $\pi\pi$
rescattering effects and are independent on any chiral one-loop input.
The most precise determinations of the charged Dalitz plot parameters come from the KLOE experiment~\cite{KLOEcharged},
see Table~\ref{tab:finalresultscharged}.
Inserting their numbers for $a$, $b$, and $d$ into Eq.~\eqref{eq:alphaisospnum}, we find
\be\label{eq:alphaisospres}
 \alpha_{\rm KLOE,NREFT}=-0.062\pm 0.003({\rm stat})_{-0.006}^{+0.004}({\rm syst})\pm 0.003({\rm \pi\pi})~,
\ee
where the statistical and systematic errors are calculated from the respective uncertainties
and their correlations in Ref.~\cite{KLOEcharged}, and the last error is the uncertainty 
inherent in our assessment of final-state interactions 
in Eq.~\eqref{eq:alphaisospnum}.
This result disagrees rather strongly with the world average of $\alpha=-0.0317\pm0.0016$~\cite{pdg}
as well as KLOE's own direct experimental finding $\alpha=-0.0301\pm0.0035_{-0.0035}^{+0.0022}$~\cite{KLOE}.

This observation seems to be at odds with a result presented in Ref.~\cite{KLOEcharged},
where a separate fit has been performed using an alternative parameterization~\cite{D'Ambrosio},
which incorporates final-state $\pi\pi$ rescattering based on a strict $\Delta I=1$ rule
and allows to extract $\alpha$ therefrom.  The result of that alternative fit is
\be
 \alpha_{\Delta I=1, {\rm exp}}=-0.038\pm 0.003({\rm stat})_{-0.008}^{+0.012}({\rm syst})~, \label{eq:alphaDAmbrosio}
\ee
and thus seems to be in very reasonable agreement with the direct determination of $\alpha$.
However, the parameterization from Ref.~\cite{D'Ambrosio} is based on chiral one-loop phases or imaginary parts,
hence leading-order rescattering with $\Order(p^2)$ $\pi\pi$ vertices.
If we reduce our rescattering formalism to that order (and also set $Q_n=Q_c$), 
we find for the coefficients in Eq.~\eqref{eq:alphaisospnum} $\zeta_1 = 0.021$, $\zeta_2 = 0.188$ instead, and as a result 
\be
  \alpha_{\Delta I=1, {\rm NREFT}}=-0.042 \pm 0.002({\rm stat})_{-0.005}^{+0.003}({\rm syst})~,
\ee
in satisfactory agreement with Eq.~\eqref{eq:alphaDAmbrosio} within errors
(which stem from the Dalitz plot input exclusively). 
We therefore understand why the rescattering formalism employed in Ref.~\cite{D'Ambrosio} leads to a seemingly
consistent result for $\alpha$; however, the large impact of higher orders in the effective range 
parameters renders this procedure unreliable.  
Employing a more precise parameterization for $\pi\pi$ final-state interactions, 
responsible for the imaginary parts necessary for the relations Eqs.~\eqref{eq:alphaineq}
and \eqref{eq:alphaisospnum},
shows that there seems to be a significant tension between the available experimental results 
for charged and neutral Dalitz plot parameters.

\subsection[Isospin-breaking corrections due to $Q_n\neq Q_c$]{Isospin-breaking corrections due to \boldmath{$Q_n\neq Q_c$}}\label{sec:RelatingDalIsobreak}

We now study isospin-breaking corrections to the above relations due to kinematic effects stemming from $Q_n\neq Q_c$.
Following the results of Sect.~\ref{sec:isospinbreaking}, all other effects are certainly
included in the uncertainties.
If we denote the charged Dalitz plot parameters as calculated in Sect.~\ref{sec:isospinlimitresults}
by $a_\text{iso}$, $b_\text{iso}$, and so forth, the ``real'' ones $a$, $b$, \ldots\ as deduced from Eq.~\eqref{eq:paramrel}
are related to the former according to
\begin{align}
a_\text{iso} &= a + \delta \left(2b-a^2\right) + \Order(\delta^2) ~, \nn
b_\text{iso} &= b + \delta \left(3f-ab\right) + \Order(\delta^2) ~, \nn
d_\text{iso} &= d + \delta \left(g-ad\right) + \Order(\delta^2) ~, \label{eq:abdshift}
\end{align}
where $\delta = Q_n/Q_c-1 \approx 0.069$, and only $f$ and $g$ do not receive corrections as long as we 
disregard Dalitz plot parameters of $\Order(\epsilon^8)$. The corrections $\propto\delta$ produce large shifts (as discussed in Sect.~\ref{sec:isospinbreaking}), 
so that one may wonder whether the relation Eq.~\eqref{eq:alphaisospnum} between charged and neutral Dalitz plot parameters may also receive large corrections.
To investigate this, we have to amend Eq.~\eqref{eq:alphaisospnum} in two respects:
\begin{enumerate}
\item
incorporate the isospin-breaking shifts due to Eq.~\eqref{eq:abdshift} in the terms $b+d-a^2/4$;
\item
improve the parameterization of ${\rm Im}(\bar a)$ to include $\Order(\epsilon^4)$ effects 
(proportional to $b$, $d$, and $a^2$ neglected before) in order to consistently 
incorporate the shifts due to Eq.~\eqref{eq:abdshift} in the contribution to $\alpha$
stemming from the imaginary part of $\bar a$.
\end{enumerate}
Although significantly more complicated in result, the manipulations are much the same as before, relying on Table~\ref{tab:param}. The improved result is of the form
\begin{align}
\alpha &=\frac{Q_n^2}{4Q_c^2}\biggl\{ b+d-\frac{a^2}{4}
- \delta\bigg[ 2a \Big(b-\frac{a^2}{4}+\frac{d}{2}\Big) -3f-g \bigg] \biggr\} \nn
& \quad - \zeta_1\Big[1+\zeta_2 a + \big(\zeta_3-\delta \zeta_2\big) a^2 + \big(\zeta_4+2\delta \zeta_2\big)  b+\zeta_5 d\Big]^2 ~, &
\zeta_1 &= 0.050 \pm 0.005 ~,   \nn
\zeta_2 &= 0.223 \pm 0.003 ~, ~~
\zeta_3 =-0.008 \pm 0.001 ~, ~~
\zeta_4 = 0.030 \pm 0.004 ~,   &
\zeta_5 &= 0.051 \pm 0.001 ~. \label{eq:alphashifted}
\end{align}
It turns out that the more refined description of ${\rm Im}(\bar a)$ in Eq.~\eqref{eq:alphashifted}
and therefore the complicated piece in the relation between charged and neutral Dalitz plot parameters
changes the result only minimally: it  shifts $\alpha$ by a mere $+0.001$.  
The term $\propto \delta$ in the first line of Eq.~\eqref{eq:alphashifted} is a bit more difficult
to evaluate, as it involves large cancellations between the various contributions. This becomes evident,
when analyzing the dependence of that term on the specific value for $f$. Varying $f$ from the experimental
result to our determination alone shifts the contribution of the $\delta$-term from $+0.002$ to $-0.002$.

For our final result we resort to the KLOE parameters again and use correlated errors, except for $g$, where there is neither a
measurement nor a determination of its correlation coefficients to be found in the literature. We simply choose to vary it independently between zero and the result of
our calculation in Table~\ref{tab:isobreak}. 
However, despite these generous variations, the total effect of these additional contributions proportional to $\delta$
is still so small that it hardly shows in the overall uncertainty.
Our final result is
\be
 \alpha_{\rm KLOE,NREFT}=-0.059\pm 0.003({\rm stat})_{-0.006}^{+0.004}({\rm syst})\pm 0.003({\rm \pi\pi})~.
\ee
The overall correction to Eq.~\eqref{eq:alphaisospres} turns out to be small and we are still left with a significant disagreement between charged and neutral channel.
Comparing the charged Dalitz plot parameters entering Eq.~\eqref{eq:alphaisospnum}, we see that
the main disagreement is due to the parameter $b$, which is strongly over-predicted in our analysis: 
we find $b=0.308\pm 0.023$ to be compared with $b_{\rm KLOE}=0.124\pm0.006\pm0.010$. 
(Of course, the NREFT results are consistent within themselves: inserting our values
for $a$, $b$, $d$ into the relation Eq.~\eqref{eq:alphaisospnum} reproduces our result for $\alpha$.)
We also mention that there is some non-negligible variation between the KLOE results for the charged Dalitz plot parameters
and several older, less precise measurements~\cite{CBcharged,Laytercharged,Gormleycharged};
a re-measurement of these quantities by some of the modern high-precision experiments would therefore
be very welcome.

\FIGURE[t]{
 \centering
 \includegraphics[width=0.7\linewidth]{abplot.eps}
 \caption{Allowed range for charged Dalitz plot parameters $a$ and $b$ with fixed $\alpha$ and $d$. Solid grey area: allowed range according to Eq.~\eqref{eq:alphaisospnum}. Hatched grey area: allowed range using $\zeta_1=0$.}
\label{fig:abplot}
}
The relation between $\alpha$ and the charged Dalitz plot parameters is further illustrated in Fig.~\ref{fig:abplot}. 
Due to the smallness of the higher-order corrections in Eq.~\eqref{eq:alphashifted}, it suffices to use the simplified 
representation Eq.~\eqref{eq:alphaisospnum}. 
As $\alpha=0.0317\pm0.0016$ is experimentally agreed upon to very high precision, 
and as our result for $d$ agrees well with the KLOE determination $d=0.057\pm0.006_{-0.016}^{+0.007}$,
we may take these two experimental results for granted, such that Eq.~\eqref{eq:alphaisospnum}
provides a relation between $a$ and $b$.
This constraint in the $a\!-\!b$ plane is shown in Fig.~\ref{fig:abplot}. 
The solid grey area shows the allowed range for $b$ as a function of $a$ according to Eq.~\eqref{eq:alphaisospnum}, 
whereas the hatched grey area shows the same relation for $\zeta_1=0$, i.e.\ fully neglecting the imaginary
part of the amplitude, or ${\rm Im}(\bar a)=0$. 
While the NREFT prediction for $a$ and $b$ falls nicely into the allowed band
(the agreement here looks even better than in the direct comparison to $\alpha$ 
as the band also reflects the experimental error in $d$), 
the KLOE determination of both is consistent with a vanishing imaginary part.
In our framework these latter values cannot be brought into agreement with a consistent implementation
of final-state interactions.

Comparing our calculation to the dispersion-theoretical analysis of Ref.~\cite{CLPeta3pi}, 
there are indications that the discrepancy we find may be slightly over-predicted:
in the terminology of the iterative solution determined there, our two-loop calculation
cannot be expected to be better than the second iteration of the dispersive amplitude.
Ref.~\cite{CLPeta3pi} shows that while the real part of the amplitude has converged to 
the final result almost perfectly, there are still non-negligible corrections in the imaginary part
beyond that, i.e.\ in terms of our representation at (irreducible) three loops and higher.
Whether those corrections in the imaginary part that precisely constitute the additional terms
in the relations Eqs.~\eqref{eq:alphaisospnum}, \eqref{eq:alphashifted} are sufficient 
to reduce the discrepancy between charged and neutral Dalitz plot parameter measurements remains to be seen.

\section{Partial widths and the ratio \boldmath{$r$}}
\label{sec:widths}
In this work, we have concentrated almost exclusively on the energy dependence of the
two $\eta\to3\pi$ Dalitz plot distributions, mainly as encoded in the Dalitz plot parameters.
It is rather obvious in particular from Table~\ref{tab:DalitzT12u} that the overall
\emph{normalization} of the amplitudes is not improved in our formalism compared to what 
we match our parameters to, here ChPT at $\Order(p^4)$; indeed, the overall rates are 
even slightly smaller.
To be concrete, integrating the Dalitz plot distributions of our amplitudes including all
isospin-breaking effects (corresponding to the last column of Table~\ref{tab:isobreak}), we find
\begin{align}
\Gamma(\eta\to3\pi^0) &= \big[ 201 \pm 3(\pi\pi) \pm 6(\Delta_{\tilde\N})\big]\,{\rm eV} ~, \nn
\Gamma(\eta\to\pi^+\pi^-\pi^0)&= \big[ 144\pm2(\pi\pi)\big]\,{\rm eV} ~. \label{eq:widths}
\end{align}
Several remarks are in order here.  
First, as we have pointed out earlier, for our normalization we use $\mathcal{Q}=24.2$ as given 
by means of Dashen's theorem, which leads to a very small width.  Changing the value to $\mathcal{Q}= 22.3$~\cite{CLPeta3pi}, 
say, immediately increases the widths by nearly 40\%.  
Second, Ref.~\cite{BGeta3pi} finds that next-to-next-to-leading order chiral corrections increase
the width by nearly 70\%, thus bringing it a lot closer to the experimental value of about 
$\Gamma(\eta\to\pi^+\pi^-\pi^0) \approx (296\pm16)\,{\rm eV}$~\cite{pdg}.
We wish to emphasize once more that this failure to reproduce the chiral enhancements
in the width in the non-relativistic framework does not invalidate our predictions for the Dalitz 
plot parameters: the power counting argument of Sect.~\ref{subsec:powcounteta3pi} explains why 
we catch the important rescattering effects in particular for the higher-order energy dependence, 
but not in the overall normalization.
The $\eta\to3\pi$ tree-level coupling constants that receive sizeable quark-mass renormalization effects
nicely factor out of the complete (tree plus loop) amplitudes and play no role in the calculation
of the Dalitz plot parameters.
As a third remark, the errors shown in Eq.~\eqref{eq:widths} do not at all reflect these uncertainties
from our matching procedure, but purely the one due to $\pi\pi$ final-state interactions (determined
as in the previous sections), and in the case of $\Gamma(\eta\to3\pi^0)$ due to the uncertainty
in $\Delta_{\tilde\N}$, see Eq.~\eqref{eq:Dalrelcorr}.

Despite all the above-mentioned deficits in a calculation of the decay widths, the ratio of
neutral-to-charged partial widths $r$ should be predicted much more reliably, as the normalization
of the amplitude (in the isospin limit) drops out.  In particular, here we may expect a somewhat
heightened importance of isospin-breaking corrections~\cite{DKMeta3pi}.  We find
\be \label{eq:r}
r = \frac{\Gamma(\eta\to3\pi^0)}{\Gamma(\eta\to\pi^+\pi^-\pi^0)} = 1.40 \pm 0.01(\pi\pi) \pm 0.04(\Delta_{\tilde\N}) ~,
\ee
in agreement with the experimental finding $r = 1.43 \pm 0.02$~\cite{pdg}.  
We note that the dependence on $\pi\pi$ rescattering in Eq.~\eqref{eq:r} is very small, 
our error is dominated by the 1.5\% uncertainty in $\Delta_{\tilde\N}$.
Equation~\eqref{eq:r} is extremely accurately reproduced by just integrating the phenomenological 
Dalitz plot distribution, with our values for the Dalitz plot parameters from Table~\ref{tab:isobreak} (last column)
instead of the exact amplitudes: obviously $r$ is affected by cusps in the neutral channel
or yet-higher-order Dalitz plot parameters at or below the permille level.
We can therefore easily derive the dependence of $r$ on the parameters $a$, $b$, $d$, \ldots, 
making use of the relation of $\alpha$ to these in Eq.~\eqref{eq:alphaineq} and neglecting
pieces that affect $r$ at the permille level (e.g.\ the terms $\propto\beta,\,\gamma$
in the neutral rate), and find
\be \label{eq:rabd}
r=1.485\,\Bigl(1-0.029\,a-0.061\,a^2+0.024\,b+0.032\,d+0.008\,f-0.014\,g\Bigr)\left(1+\frac{2\Delta_{\tilde\N}}{\N_n}\right)~.
\ee
Errors on this result are to be taken from Eq.~\eqref{eq:r}. 
The various numerical coefficients are given by $\pi\pi$ phase shifts and phase space integration only. 
This demonstrates to very good approximation that $r$ does not depend on the normalization and thus 
possibly sizeable quark mass renormalization effects.

\section{Summary and conclusion}
\label{sec:summary}
In this article we have analyzed rescattering effects in $\eta\to3\pi$ decays by means of the 
modified non-relativistic effective field theory framework.
The main findings of our investigation can be summarized as follows:
\begin{enumerate}
\item 
NREFT provides a  simple and transparent representation of the amplitude to two loops, including higher-order
isospin breaking. 
In order to estimate higher-order loop effects we have furthermore applied a simplified unitarization prescription.
The amplitude thus obtained is -- at the very least -- fully competitive with the chiral expansion at 
next-to-next-to-leading order. 
The coupling constants involved have been matched to phenomenological $\pi\pi$ scattering threshold parameters
and, in the case of the $\eta\to3\pi$ tree-level couplings, to ChPT at $\Order(p^4)$.
\item
One- and two-loop contributions to the Dalitz plot parameters are in general of the same size, 
an observation which is predicted by non-relativistic power counting arguments.
Irreducible two-loop graphs are generally suppressed, while derivative couplings at two-loop level are
essential to find the correct sign for the $\eta\to3\pi^0$ slope parameter $\alpha$.
Higher-order effects beyond two loops were shown to 
be relatively small, but not negligible.
\item
While our results for the Dalitz plot parameters are in qualitative agreement with previous dispersive results,
we can provide an explanation for the apparent failure of two-loop ChPT to reproduce $\alpha$:
the treatment of $\pi\pi$ final-state interactions is still not sufficiently accurate at that order.
We can identify one specific diagram, the double rescattering graph with $\pi\pi$ vertices 
beyond leading order, as being responsible for at least half of the discrepancy between the $\Order(p^6)$
prediction for $\alpha$ and the experimental value.  These effects are of chiral order $p^8$ and higher,
but included in the NREFT two-loop representation.
\item
Apart from normalization effects and subtleties in the definition of the center of the Dalitz plot 
in the charged decay channel, higher-order isospin-breaking corrections 
on the Dalitz plot parameters are very small.
\FIGURE[t]{
 \centering
 \includegraphics[width=0.8\linewidth]{slopes.eps}
 \caption{Comparison of values for the slope parameter $\alpha$. Top: theoretical predictions. Bottom: experimental determinations. The grey shaded area is the particle data group average~\cite{pdg}.
 \label{fig:slopes}}
}
\item
Our final result for neutral Dalitz slope parameter,
\be\label{eq:finalres}
\alpha=-0.025\pm0.005~,
\ee
is compared in Fig.~\ref{fig:slopes} to several other determinations. 
It is considerably closer to the experimental world average $\alpha=-0.0317\pm0.0016$~\cite{pdg} than previous theoretical approaches. 
Notice though that Eq.~\eqref{eq:finalres} does not take uncertainties stemming from matching to ChPT at $\order(p^4)$ 
into account, which we expect to be non-negligible.
\item
Our results for the charged Dalitz plot parameters show somewhat larger deviations
from the currently most accurate measurement by the KLOE collaboration. 
By relating charged and neutral decay channel via the $\Delta I=1$ rule we
find indications for a significant tension between the Dalitz plot parameters of both channels, 
which is solely due to final-state interactions. 
A re-measurement of the charged Dalitz plot parameters by high-precision experiments~\cite{KLOEplans,WASAplans}, 
or even preferably access to improved full Dalitz plot distributions, is thus highly desirable. 
\item
While the partial widths calculated in our framework do not improve upon the chiral one-loop prediction
we match to (due to the absence of further quark-mass renormalization effects not captured in our framework), 
we can give a value for the ratio of neutral-to-charged partial widths unaffected by this deficit, 
$r = 1.40 \pm 0.04$, where the error is dominated by isospin-breaking effects.
\end{enumerate}

Possible future improvements on the theoretical approach  include matching to $\order(p^6)$ ChPT in order to constrain the tree-level Dalitz plot couplings more tightly. 
Furthermore, it will be extremely useful to match the non-relativistic representation to the upcoming dispersive analysis~\cite{CLPeta3pi} in order to obtain a reliable description of the whole physical Dalitz plot:
in this way one can include elastic $\pi\pi$ rescattering to all orders, 
and at the same time implement in particular non-analytic effects (cusps) at or near the boundaries of the Dalitz plot
due to isospin-breaking up to next-to-next-to-leading order.
This combination should then also provide the best-possible representation of the decay amplitude
for a precision extraction of the quark mass ratio $\mathcal{Q}$.

\section*{Acknowledgements}
We would like to thank Gilberto Colangelo, J\"urg Gasser, Andrzej Kup\'s\'c, Stefan Lanz, and Akaki Rusetsky for stimulating discussions,
and J\"urg Gasser for numerous useful remarks on the manuscript.

\appendix

\section{Isospin-breaking corrections}

\subsection[Isospin-breaking corrections to $\pi\pi$ scattering]{Isospin-breaking corrections to \boldmath{$\pi\pi$} scattering}
\label{app:EMcorr}
To calculate isospin-breaking corrections to the matching relations Eq.~\eqref{eq:pipiMatch}, 
we expand the ChPT amplitudes for all channels with electromagnetic corrections included~\cite{KnechtUrech,KnechtNehme}
around threshold. 
These contain virtual-photon exchange and real-photon radiation in the form of bremsstrahlung. 
For the definition of a reasonable (regular) threshold expansion, 
at first the divergent Coulomb pole contribution has to be subtracted.
As the Coulomb pole emerges equally in the vertex correction diagram in both 
the NREFT and the ChPT calculation due to the same infrared properties of both theories, 
in a matching between them, this part drops out anyway.
The  determination of the scattering lengths is then relatively straightforward and has already been performed in the above references. Note, however, that in contrast to Refs.~\cite{KnechtUrech,KnechtNehme} we expand around an isospin limit defined in terms 
of the \emph{charged} pion mass. Thus, non-analytic terms $\propto\sqrt{\Delta_\pi}$ arise in the expansion of the $\pi^0\pi^0\to\pi^0\pi^0$ channel, which are due to a cusp structure at the charged pion threshold and cancel the corresponding 
contribution in the expansion of $J_{+-}(s)$ in Eq.~\eqref{eq:nonrelscatt}, once the correct matching is performed. We display the corrections in the form $C_i=\bar C_i+\Delta C_i$, 
where $\bar C_i$ denotes the corresponding coupling in the isospin limit
(the $\pi^+\pi^+\to\pi^+\pi^+$ channel is not needed in the present analysis, we just 
give it for completeness). For the combinations of S-wave scattering lengths, we find
\begin{align}
\Delta C_{00}&=\frac{\mpc^2}{F_\pi^2}\Bigl\{-\frac{\Delta_\pi}{\mpc^2}+\frac{e^2}{32 \pi^2}\K^{00}+\frac{\Delta_\pi}{32 \pi^2  F_\pi^2}\bigl(13-16 \bar{l}_1-32 \bar{l}_2+6 \bar{l}_3-4 \bar{l}_4\bigr)\Bigr\}~,\nn
\Delta C_{x}&=\frac{\mpc^2}{F_\pi^2}\Bigl\{-\frac{\Delta_\pi}{\mpc^2}+\frac{e^2}{32 \pi^2}\bigl(30-3 \K_1^{\pm 0}+\K_2^{\pm 0}\bigr)-\frac{\Delta_\pi}{96 \pi^2  F_\pi^2}\bigl(23+8 \bar{l}_1+6 \bar{l}_3+12 \bar{l}_4\bigr)\Bigr\}~,\nn
\Delta C_{+0}&=\frac{\mpc^2}{F_\pi^2}\Bigl\{\frac{\Delta_\pi}{\mpc^2}-\frac{e^2}{32 \pi^2}\bigl(2+\K_1^{\pm 0}+\K_2^{\pm 0}\bigr)+\frac{\Delta_\pi}{96 \pi^2  F_\pi^2}\bigl(3-8 \bar{l}_1-16 \bar{l}_2+6 \bar{l}_3+12 \bar{l}_4\bigr)\Bigr\}~,\nn
\Delta C_{+-}&=\frac{\mpc^2}{F_\pi^2}\Bigl\{\frac{2\Delta_\pi}{\mpc^2}-\frac{e^2}{16 \pi^2}\bigl(24-\K^{+-}\bigr)+\frac{\Delta_\pi}{8 \pi ^2 F_\pi^2}\bigl(2+\bar{l}_3+2\bar{l}_4\bigr)\Bigr\}~,\nn
\Delta C_{++}&=\frac{\mpc^2}{F_\pi^2}\Bigl\{\frac{2\Delta_\pi}{\mpc^2}-\frac{e^2}{16 \pi^2}\bigl(20-\K^{++}\bigr)+\frac{\Delta_\pi}{16 \pi ^2 F_\pi^2}\bigl(3+2\bar{l}_3+4 \bar{l}_4\bigr)\Bigr\}~.
\end{align}
We note that $\Delta C_{00}$ is indeed free of non-analytic terms in $\Delta_\pi$:
the analytic structure of ChPT and the non-relativistic representation near threshold is the same, as it must.

The definition of effective ranges and P-wave scattering lengths is not \emph{a priori} clear, since one has to deal with infrared divergences in the ChPT amplitudes. In calculations of, say,  cross sections these divergences, which arise
from virtual-photon corrections, cancel with corresponding divergences from real-photon radiation (bremsstrahlung). 
However, when matching the non-relativistic framework to ChPT, the explicit inclusion of bremsstrahlung is not necessary, since the virtual-photon diagrams exhibit the same infrared behavior and thus contain the same divergences (see Ref.~\cite{RadK3pi}). On a rather technical note, the infrared divergences were calculated in dimensional regularization in Ref.~\cite{RadK3pi}, 
while the ChPT calculations~\cite{KnechtUrech,KnechtNehme} use a finite photon mass $m_\gamma$ as infrared regulator.
The transition between both regularization schemes can be made by replacing $\log({m_\gamma^2}/{\mpc^2})\to-32\pi^2\lambda_{\rm IR}-1$. The infrared divergences then cancel, rendering the matching relations finite. 
We wish to emphasize that the physical reason for this cancellation
is again the identical infrared behavior of both theories.

For the S-wave effective ranges, defining in analogy with the above $D_i = \bar D_i + \Delta D_i$, 
we find the following corrections to the matching relations:
\begin{align}
\Delta D_{00}&=\frac{1}{F_\pi^2}\Bigl\{\frac{\Delta_\pi}{48 \pi^2  F_\pi^2}\bigl(35-8 \bar{l}_1-16 \bar{l}_2\bigr)\Bigr\}~,\nn
\Delta D_{x }&=\frac{1}{F_\pi^2}\Bigl\{\frac{e^2}{96 \pi^2}\bigl(59-3 \K_1^{\pm 0}\bigr)+\frac{ \Delta_\pi}{120 \pi^2  F_\pi^2}\bigl(18-5 \bar{l}_1\bigr)\Bigr\}~,\nn
\Delta D_{+0}&=\frac{1}{F_\pi^2}\Bigl\{-\frac{e^2}{192 \pi^2}\bigl(1+3 \K_1^{\pm 0}\bigr)+\frac{\Delta_\pi}{192 \pi^2  F_\pi^2}\bigl(21-4 \bar{l}_1-12 \bar{l}_2\bigr)\Bigr\}~,\nn
\Delta D_{+-}&=\frac{1}{F_\pi^2}\Bigl\{-\frac{e^2}{1152\pi^2}\Bigl(764-9 (\K^{+-}- \K^{++})\Bigr)-\frac{109\Delta_\pi}{384\pi ^2 F_\pi^2}\Bigr\}~,\nn
\Delta D_{++}&=\frac{1}{F_\pi^2}\Bigl\{-\frac{e^2}{576 \pi^2}\Bigl(676+9 (\K^{+-}- \K^{++})\Bigr)+\frac{61\Delta_\pi}{192\pi ^2 F_\pi^4}\Bigr\}~,
\end{align}
while for the two P-wave scattering lengths, we have (with $E_i = \bar E_i + \Delta E_i$)
\begin{align}
\Delta E_{+0}&=\frac{1}{F_\pi^2}\Bigl\{\frac{e^2}{64 \pi^2}\bigl(1+3 \K_1^{\pm 0}\bigr)-\frac{\Delta_\pi}{192 \pi^2  F_\pi^2}\bigl(19-12 \bar{l}_1+12 \bar{l}_2\bigr)\Bigr\}~,\nn
\Delta E_{+-}&=\frac{1}{F_\pi^2}\Bigl\{\frac{3e^2}{128\pi^2}\Bigl(-28+\K^{+-}- \K^{++}\Bigr)-\frac{93 \Delta_\pi}{128\pi^2  F_\pi^2}\Bigr\}~,
\end{align}
where the following abbreviations have been used for combinations of electromagnetic SU(2) low-energy constants $\bar k_i$ and $ Z = {\Delta_\pi}/({2e^2 F_{\pi }^2})$: 
\begin{align}
  \K^{00}&=\Bigl(3 + \frac{4Z}{9}\Bigr)\bar k_1 - \frac{40Z}{9} \bar k_2 - 3\bar k_3 - 4Z\bar k_4~,\nn
  \K_1^{\pm 0}&=\Bigl(3 + \frac{4Z}{9}\Bigr) \bar k_1 + \frac{32Z}{9}\bar k_2 + 3\bar k_3 + 4Z\bar k_4~,\nn
  \K_2^{\pm 0}&=8Zk_2 + 3\bar k_3 + 4Z\bar k_4 - 2(1 + 8Z)\bar k_6 - (1-8Z)\bar k_8~,\nn
  \K^{+-}&=\Bigl(3 + \frac{4Z}{9}\Bigr)\bar k_1 - \frac{40Z}{9} \bar k_2 - 9\bar k_3 + 4Z\bar k_4 + 4(1 + 8Z)\bar k_6 + 2(1-8Z)\bar k_8~,\nn
  \K^{+-}-\K^{++}&=2\Bigl(3+\frac{4 Z}{9}\Bigr) \bar k_1+\frac{208 Z \bar k_2}{9}-18 \bar k_3+24 Z \bar k_4~.
\end{align}
We refrain from calculating corrections to the shape parameters, since their intrinsic, isospin-symmetric error is much larger than what can be expected from isospin breaking.

For the numerical evaluation we express the low-energy constants $\bar l_1$ and $\bar l_2$ in terms of $\pi\pi$ 
D-wave scattering lengths~\cite{GLoneloop}, for which we use the numerical values~\cite{CGL}
\be
a_2^0=1.75\pm0.03\times 10^{-3}\mpc^{-4}~,\qquad
a_2^2=0.170\pm0.013\times10^{-3}\mpc^{-4}~.
\ee
This way $a_2^0$ and $a_2^2$ can be independently varied according to their uncertainty, whereas $\bar l_1$ and $\bar l_2$ are correlated. 
For $\bar l_3$ we propose
$
\bar l_3=3.1\pm0.5 
$
as a sensible mean value from lattice simulations (see Ref.~\cite{Necco} for individual results of the various groups). 
The constant $\bar l_4$ is extracted from the scalar radius of the pion~\cite{CGL},
$
\bar l_4=4.4\pm0.2\,.
$

For the electromagnetic SU(2) low-energy constants $k_i^r$ we use the values given in Ref.~\cite{Haefeli}. 
The authors of this work have matched the two-flavor low-energy constants to their SU(3) counterparts, using
numerical estimates from Refs.~\cite{KLECs,KLECs2}. 
We convert the values $k_i^r$ given at the mass of the $\rho$, $M_\rho=0.77$~GeV, in Ref.~\cite{Haefeli}
to scale-independent constants according to the standard prescription,
\be
  \bar k_i = \frac{32\pi^2}{\sigma_i}k_i^r(M_\rho)-\log{\frac{\mpc^2}{M_\rho^2}}~,
\ee
where the $\sigma_i$ are the corresponding $\beta$-functions to be found in Ref.~\cite{KnechtUrech}.
Numerically this results in
\be
\bar k_1=1.66~, \quad  \bar k_2=4.08~, \quad \bar k_3=2.28~, \quad \bar k_4=3.69~, \quad
\bar k_6=4.08~, \quad  \bar k_8=4.06~.
\ee
The uncertainties on the $k_i^r$ are estimated analogously to Ref.~\cite{DKMeta3pi} by their logarithmic scale variation,
\be \label{eq:emerror}
  k_i^r\to k_i^r\pm\frac{\sigma_i}{16\pi^2}~,
\ee
which for the $\bar k_i$ translates to
$
  \bar k_i\to \bar k_i \pm 2 \,.
$
The errors on the quantities $\K^{00},\K_i^{\pm 0},\K^{+-},\K^{++}$ are then calculated in a correlated fashion 
(i.e.\ $+2$ or $-2$ for \emph{all} $\bar k_i$).

The numerical corrections are displayed in Table~\ref{tab:scattcorr}.
\TABLE[t]{
\centering
\renewcommand{\arraystretch}{1.2}
\begin{tabular}{c rcl rcl rcl}
 \toprule
channel	&$\Delta C_i/C_i$&\tbk$\times$&\tbk$10^{-2}$ 	& $\Delta D_i/D_i$&\tbk$\times$&\tbk$10^{-2}$	& $\Delta E_i/E_i$&\tbk$\times$&\tbk$10^{-2}$\\
\midrule
$00$	&$-7.3$		 &\tpm&\tbk$0.2$		& $-3.3$	  &\tpm&\tbk$0.4$ 		& 		  &\tbk--&\tbk \\
$x$	&$2.5$		 &\tpm&\tbk$0.6$ 		& $0.1$		  &\tpm&\tbk$0.6$ 		&		  &\tbk--&\tbk \\
$+0$ 	&$-5.2$		 &\tpm&\tbk$0.8$ 		& $1.9$		  &\tpm&\tbk$0.8$ 		&$0.4$	  	  &\tpm&\tbk$0.6$ \\
$+-$	&$6.1$		 &\tpm&\tbk$0.5$ 		& $-0.2$	  &\tpm&\tbk$0.3$ 		&$0.5$	  	  &\tpm&\tbk$0.4$ \\
\bottomrule
\end{tabular}
\renewcommand{\arraystretch}{1.0}
\caption{Corrections to the matching relations relative to the phenomenological values.\label{tab:scattcorr}}
}
We find that the corrections at one-loop order are very small. The main contributions to the $C_i$ stem from the tree-level correction factor.

\subsection[Corrections to the $\Delta I=1$ rule]{Corrections to the \boldmath{$\Delta I=1$} rule}
\label{app:isosprelcorr}
At leading order $p^2$ in ChPT and up to next-to-leading order in the isospin-breaking parameters 
$m_u-m_d$ and $e^2$, the amplitudes for the charged and the neutral decay were already quoted in Eq.~\eqref{eq:LOChPTamps}.
We also hinted at the fact that in order to define the deviations from the $\Delta I =1$ relation,
Eq.~\eqref{eq:Dalrelcorr}, it is useful to expand the decay amplitudes for \emph{both} channels
around the point $s_3=s_n$, $s_1=s_2$ as shown in Eq.~\eqref{eq:ampsqS1S2S3}, so that $\Delta_{\tilde \N}$ is going to be of chiral order $p^4$.

The decay amplitudes at $\Order(p^4)$ in ChPT and at $\order(m_d-m_u,e^2,(m_d-m_u)e^2)$ in isospin breaking 
are given explicitly in Ref.~\cite{DKMeta3pi}.  
With minimal modifications they can be shown to be also valid up-to-and-including $\order((m_d-m_u)^2)$, i.e.\ only numerically tiny terms of $\order(e^4)$ are potentially neglected at second order in isospin breaking. 
In order to match the expanded ChPT amplitude of Ref.~\cite{DKMeta3pi} to the polynomial part 
of the NREFT representation, the following steps have to be taken into account:
\begin{enumerate}
\item
The (non-analytic) imaginary  parts due to pion loops in chiral and NREFT amplitude are identical and
drop out in the matching relation.
\item
For the radiative corrections due to real and virtual photons, we have to match the result
of Ref.~\cite{DKMeta3pi} to an analogous NREFT representation as in Ref.~\cite{RadK3pi}.
As a result, the Coulomb pole and phase have to be subtracted from the chiral representation,
as well as the bremsstrahlung contributions. As in the case of radiative corrections to $\pi\pi$ scattering
infrared divergences were regulated by introducing a finite photon mass $m_\gamma$ in Ref.~\cite{DKMeta3pi} and have to be treated
as described in Appendix~\ref{app:EMcorr}.
\end{enumerate}

As $\Delta_{\tilde{\N}}$ is of $\Order(p^4)$, it is convenient to factor out the neutral normalization at leading order 
and quote the result as the ratio $\Delta_{\tilde{\N}}/\N_n$ below.
We find
\begin{align}\label{eq:deltatildeNrelative}
  \frac{\Delta_{\tilde{\N}}}{\N_n}&=2e^2\bigg\{\frac{1-3\rho}{3\rho}G(s_n)+\frac{1}{2}\bar{J}_{\pi\pi}(s_n)+\frac{3}{32\pi^2}\left(\log\frac{\mpc^2}{\mu^2}-1\right)-\frac{1+\rho}{1-\rho}(2K_3^r-K_4^r)\nn
      &\qquad+\frac{8K_6^r}{3(1-\rho)}-\frac{4(3-\rho)}{1-\rho}(K_{10}^r+K_{11}^r)\bigg\}\nn
      &\quad+\frac{\Delta_\pi}{3(1-\rho)F_\pi^2}\bigg\{\frac{29-111\rho-9\rho^2+27\rho^3}{8(1+3\rho)}\bar{J}_{KK}(s_n)-32L_3^r\nn
      &\qquad+\frac{3\rho(1+22\rho+9\rho^2)}{(1-9\rho)(1+3\rho)}\bar{J}_{\eta\pi}(s_n)-\frac{7+21\rho-495\rho^2+243\rho^3}{(1-9\rho)(1+3\rho)}\bar{J}_{\pi\pi}(s_n)\nn
      &\qquad+\frac{8(3-\rho)}{1-\rho}\frac{F_\pi^2}{\meta^2}\Delta_F+\frac{1}{16\pi^2}\bigg[6(3-2\rho)\log\frac{\mpc^2}{\mu^2}+\frac{2(1+2\rho-\rho^2)}{1-\rho}\log\frac{3+\rho}{4\rho}\nn
        &\qquad\quad-\frac{3(3-26\rho-\rho^2)}{(1-9\rho)(1-\rho)}\log\rho+\frac{53-357\rho+351\rho^2+81\rho^3}{4(1-9\rho)}\bigg]\bigg\}+\order\left(e^2p^2\right)~,
\\
\label{eq:deltatildealpha}
  \Delta_{\tilde{\alpha}}&=\frac{3e^2}{(1-9\rho)^2(1-\rho)(1+3\rho)^2\meta^4}\bigg\{12\rho(1+63\rho^2)G(s_n)\nn
      &\qquad-\frac{1-14\rho-138\rho^2+234\rho^3-1107\rho^4}{1-9\rho}\bar{J}_{\pi\pi}(s_n)\nn
      &\qquad+\frac{7-102\rho-504\rho^2+1926\rho^3-3375\rho^4}{32\pi^2(1-9\rho)}\bigg\}\nn
      &+\frac{3\Delta_\pi}{(1-\rho)(1+3\rho)^2F_\pi^2\meta^4}\bigg\{\frac{3(3+\rho)(957-5240\rho-1398\rho^2-288\rho^3+81\rho^4)}{4096(1+3\rho)}\bar{J}_{KK}(s_n)\nn
      &\qquad+\frac{\rho^2(221-3612\rho+32022\rho^2-32076\rho^3-2187\rho^4)}{8(1-9\rho)^3(1+3\rho)}\bar{J}_{\eta\pi}(s_n)\nn
      &\qquad-\frac{3\rho(3-124\rho+1794\rho^2-7596\rho^3+9315\rho^4)}{(1-9\rho)^3(1+3\rho)}\bar{J}_{\pi\pi}(s_n)\nn
      &\qquad+\frac{1}{32768\pi^2(1-9\rho)^3}\Bigl(243-39737\rho+540471\rho^2-729333\rho^3\nn
      &\quad\qquad+3630825\rho^4-1810107\rho^5+85293\rho^6+59049\rho^7\Bigr)\nn
      &\qquad-\frac{\rho^2(37+237\rho-2025\rho^2+3159\rho^3)}{128\pi^2(1-9\rho)^3(1-\rho)}\log\rho\bigg\}+\order\left(e^2p^{-2}\right)~,
\end{align}
neglecting even higher-order terms in the isospin-breaking parameters $e^2$ and $m_d-m_u$.
Here, we have used $\Delta_F=F_K/F_\pi-1$ (cf.\ Ref.~\cite{GLNPB250}) and the loop functions
\begin{align} \label{eq:Jbar}
  G(s) &=\frac{1-\sigma_\pi^2}{64\pi^2\sigma_\pi}\bigg\{\operatorname{Li}\Big(\frac{1-\sigma_\pi}{1+\sigma_\pi}\Big)
-\operatorname{Li}\Big(\frac{1+\sigma_\pi}{1-\sigma_\pi}\Big)+\log\frac{1+\sigma_\pi}{1-\sigma_\pi}\bigg\}~, \quad
\operatorname{Li}(z)=\int_{1}^{z}\frac{\log t}{1-t}dt ~, \nn
  \bar{J}_{\pi\pi}(s) &= \frac{1}{8\pi^2}\bigg\{1-\frac{\sigma_\pi}{2}\log\frac{1+\sigma_\pi}{1-\sigma_\pi}\bigg\}~,\qquad
  \bar{J}_{KK}(s)=\frac{1}{8\pi^2}\big\{1-\sigma_K\operatorname{arccot}\sigma_K\big\}~,\nn
  \bar{J}_{\eta\pi}(s)&=\frac{1}{32\pi^2}\bigg\{2+\log\rho\bigg(\frac{\meta^2-\mpc^2}{s}-\frac{1+\rho}{1-\rho}\bigg)-\frac{\nu}{s}\log\frac{s-\meta^2-\mpc^2+\nu}{s-\meta^2-\mpc^2+\nu}\bigg\}~, \nn
  \sigma_\pi&=\sqrt{1-\frac{4\mpc^2}{s}}~, \quad
  \sigma_K=\sqrt{\frac{4M_K^2}{s}-1}~, \quad 
  \nu=\lambda^{1/2}(\meta^2,\mpc^2,s)~, \quad \rho=\frac{\mpc^2}{\meta^2}~.
\end{align}
$G(s)$ is the real part of the triangle loop function for the photon exchange between two charged pions 
(rescaled by a factor of $\mpc^2$) with the Coulomb pole subtracted, involving Spence's function $\operatorname{Li}(z)$,
and $\bar{J}_{ab}(s)$ are the usual finite and scale-independent parts of the corresponding two-meson loop functions. 
For the definition of the (renormalized) strong and electromagnetic SU(3) low-energy constants $L_3^r$ and $K_i^r$ in terms of chiral Lagrangians (not to be confused with the tree-level couplings $L_i$, $K_i$ of the non-relativistic theory), 
see Refs.~\cite{GLNPB250,Urech}.
We have made extensive use of the Gell-Mann--Okubo relation to simplify the results Eqs.~\eqref{eq:deltatildeNrelative} and
\eqref{eq:deltatildealpha}.
Furthermore, the kinematic expansion to second order plus the expansion in isospin-breaking parameters
leads to derivatives up to third order of these loop functions, which have been rewritten 
in terms of the loop functions themselves.

Both results Eqs.~\eqref{eq:deltatildeNrelative} and \eqref{eq:deltatildealpha} 
are divergence-free and independent of the scale $\mu$.
The scale-independence of $\Delta_{\tilde{\alpha}}$ is explicitly seen, 
that of $\Delta_{\tilde\N}$ can be found by using the scale variation of 
the electromagnetic constants $K_i^r$ as given in Ref.~\cite{Urech}.
Both corrections turn out to be completely of electromagnetic origin.
For the numerical evaluation we make use of the same estimates and variations of the low-energy
constants as described in Appendix~\ref{app:EMcorr} and explained in more detail in Ref.~\cite{DKMeta3pi};
their uncertainties completely dominate the error on ${\Delta_{\tilde \N}}/{\N_n}$.
As $\Delta_{\tilde\alpha}$ is free of low-energy constants at this order, it is
a pure loop effect and a prediction in terms of well-known parameters.
Here we quote an uncertainty solely due to the use of the Gell-Mann--Okubo relation for the masses,
using either the mass of the $\eta$ directly, or the same expressed in terms of pion and kaon masses.
We consider the error thus obtained rather underestimated.  
In total, we find
\be
  \frac{\Delta_{\tilde \N}}{\N_n}=(-0.7\pm1.5)\%~,\qquad \Delta_{\tilde\alpha}=0.035\pm0.003~{\rm GeV}^{-4}~.
\ee

\section{NREFT representation including isospin breaking}
\label{app:NREFTrep}

\subsection[$\eta\to 3\pi$ amplitudes up to two loops]{\boldmath{$\eta\to 3\pi$} amplitudes up to two loops}
\label{app:NREFT2loop}
For the representation of the $\eta\to3\pi$ decay amplitudes at one-loop order, we find (see also Ref.~\cite{KL3pi})
\begin{align}\label{eq:1-loop}
\M_n^{\text{1-loop}}(s_1,&s_2,s_3)=\Bigl\{C_{00}(s_1)K(s_1)J_{00}(s_1)+2C_x(s_1)L(s_1)J_{+-}(s_1)+(s_1 \leftrightarrow s_2) \nn
&\qquad\qquad+ (s_1 \leftrightarrow s_3)\Bigr\}~,\nn
\M_c^{\text{1-loop}}(s_1,&s_2,s_3)=C_x(s_3)K(s_3)J_{00}(s_3)+2C_{+-}(s_3)L(s_3)J_{+-}(s_3)\nn
&\qquad+\Bigl\{\Bigl[2C_{+0}(s_1)L'(s_1)-\tilde{E}_{+0}(s_1,s_2,s_3)\tilde{L}(s_1)\Bigr]J_{+0}(s_1)+(s_1 \leftrightarrow s_2)\Bigr\}~.
\end{align}
The topologies at two-loop order are shown in Fig.~\ref{fig:twolooptopologies}. For the two-loop amplitudes at the order discussed in Sect.~\ref{subsec:powcounteta3pi}, we obtain
\begin{align}\label{eq:2-loop}
\M_n^{\words{2-loops}} &=\Bigl\{\M_n^A(s_1,s_2,s_3) +\M_n^B(s_1,s_2,s_3) + (s_1\leftrightarrow s_2) + (s_1\leftrightarrow s_3)\Bigr\}~,\nn
\M_c^{\words{2-loops}} &=\M_c^A(s_1,s_2,s_3)+\M_n^B(s_1,s_2,s_3)~,
\end{align}
where
\begin{align}
\M_n^A
%(1)
&=2 K_0 C_{00}(\tilde{s}_1^{00}) C_{00}(s_1)F_0(\mpn,\mpn,\mpn,\mpn,s_1)\nn
&\quad-4\meta K_0 D_{00} C_{00}(s_1) \frac{\vec{Q}_1^2}{Q_1^0} F_0^{(1)}(\mpn,\mpn,\mpn,\mpn,s_1)\nn
%(2)
&+8\Bigl[ L'''_{+0}(s_1) C_{+0}(\tilde{s}_1^{+-}) \nn
&\qquad- \frac{\Delta_\pi}{4\mpn} \Bigl(\frac{s_1+2\vec{Q}_1^2}{2Q_1^0}-p_1^0\Bigr) L_0 E_{+0}\Bigr] C_x(s_1) F_0(\mpn,\mpc,\mpc,\mpc,s_1)\nn
&\quad-8\Bigl[\frac{1}{2}L_1 C_{+0}(\tilde{s}_1^{+-})+2\meta L'''_{+0}(s_1) D_{+0} \nn
&\quad\qquad- \frac{\Delta_\pi}{4\mpn} L_0 E_{+0}\Bigr] C_x(s_1) \frac{\vec{Q}_1^2}{Q_1^0} F_0^{(1)}(\mpn,\mpc,\mpc,\mpc,s_1)\nn
&\quad+8\meta L_1 D_{+0} C_x(s_1) \frac{\vec{Q}_1^4}{(Q_1^0)^2}F_0^{(2)}(\mpn,\mpc,\mpc,\mpc,s_1)\nn
%(3)
&+4 L''(s_1) C_x(\tilde{s}_1^{00}) C_{00}(s_1) F_0(\mpc,\mpc,\mpn,\mpn,s_1)\nn
&\quad+4\Bigl[L_1 C_{x}(\tilde{s}_1^{00})-2\meta L''(s_1) D_x\Bigr] C_{00}(s_1) \frac{\vec{Q}_1^2}{Q_1^0} F_0^{(1)}(\mpc,\mpc,\mpn,\mpn,s_1)\nn
&\quad-8\meta L_1 D_x C_{00}(s_1)\frac{\vec{Q}_1^4}{(Q_1^0)^2} F_0^{(2)}(\mpc,\mpc,\mpn,\mpn,s_1)~,  \label{eq:NA} \\[5mm]
\M_n^B
&=K(s_1)C_{00}(s_1)^2J_{00}^2(s_1)+2\Bigl[L(s_1)C_x(s_1)C_{00}(s_1)+K(s_1)C_x^2(s_1)\Bigr]J_{00}(s_1)J_{+-}(s_1)\nn
&+4L(s_1)C_{+-}(s_1)C_x(s_1)J_{+-}^2(s_1) ~,\label{eq:NB} \\[5mm]
\M_c^A
%(1)
&=\biggl\{4\Bigl[L'''_{+0}(s_1^+)C_{+0}(\tilde{s}_1^{+0})\Bigl(C_{+0}(s_1)-E_{+0}^{+}(s_1,s_2,s_3)\Bigr) \nn
&\qquad + \frac{\Delta_\pi}{4\mpn} \Bigl(\frac{s_1+2\vec{Q}_1^2-\Delta_\pi}{2Q_1^0}-p_1^0\Bigr) L_0 E_{+0}C_{+0}(s_1)\Bigr] F_+(\mpc,\mpn,\mpc,\mpn,s_1) \nn
&\quad - 4\Bigl[\Bigl(\frac{1}{2}L_1 C_{+0}(\tilde{s}_1^{+0}) + 2\meta L_{+0}'''(s_1^+)D_{+0}\Bigr)\Bigl(C_{+0}(s_1)-E_{+0}^{+}(s_1,s_2,s_3)\Bigr)\frac{\vec{Q}_1^2}{Q_1^0} \nn
&\qquad + \frac{\Delta_\pi}{4\mpn} L_0 E_{+0}C_{+0}(s_1)\frac{\vec{Q}_1^2}{Q_1^0} \nn
&\qquad- 2L'''_{+0}(s_1^+)C_{+0}(\tilde{s}_1^{+0})E_{+0}(s_1,s_2,s_3) \Bigr] F_+^{(1)}(\mpc,\mpn,\mpc,\mpn,s_1) \nn
&\quad + 4\Bigl[\meta L_1D_{+0}\Bigl(C_{+0}(s_1)-E_{+0}^+(s_1,s_2,s_3)\Bigr)\frac{\vec{Q}_1^4}{(Q_1^0)^2}-\Bigl(L_1 C_{+0}(\tilde{s}_1^{+0}) \nn
&\qquad\qquad+4\meta L'''_{+0}(s_1^+)D_{+0}\Bigr)E_{+0}(s_1,s_2,s_3)\frac{\vec{Q}_1^2}{Q_1^0} \Bigr]F_+^{(2)}(\mpc,\mpn,\mpc,\mpn,s_1) \nn
&\quad + 8 \meta L_1 D_{+0} E_{+0}(s_1,s_2,s_3)\frac{\vec{Q}_1^4}{(Q_1^0)^2}F_+^{(3)}(\mpc,\mpn,\mpc,\mpn,s_1) \nn
%(2)
&+ 4 L''(s_1^-)C_{+-}(\tilde{s}_1^{0+})\Bigl(C_{+0}(s_1)+E_{+0}^{-}(s_1,s_2,s_3)\Bigr) F_+(\mpc,\mpc,\mpn,\mpc,s_1) \nn
&\quad + 4 \Bigl[\Bigl(L_1 C_{+-}(\tilde{s}_1^{0+})-2\meta L''(s_1^-)D_{+-}\Bigr)\Bigl(C_{+0}(s_1)+E_{+0}^{-}(s_1,s_2,s_3)\Bigr)\frac{\vec{Q}_1^2}{Q_1^0} \nn
&\qquad - 2L''(s_1^-)C_{+-}(\tilde{s}_1^{0+})E_{+0}(s_1,s_2,s_3)\Bigr] F_+^{(1)}(\mpc,\mpc,\mpn,\mpc,s_1) \nn
&\quad - 8\Bigl[\meta L_1D_{+-}\Bigl(C_{+0}(s_1)+E_{+0}^-(s_1,s_2,s_3)\Bigr)\frac{\vec{Q}_1^4}{(Q_1^0)^2} + \Bigl(L_1 C_{+-}(\tilde{s}_1^{0+})\nn
&\qquad\qquad -2\meta L''(s_1^-)D_{+-}\Bigr)E_{+0}(s_1,s_2,s_3)\frac{\vec{Q}_1^2}{Q_1^0} \Bigr]F_+^{(2)}(\mpc,\mpc,\mpn,\mpc,s_1) \nn
&\quad + 16 \meta L_1D_{+-}E_{+0}(s_1,s_2,s_3) \frac{\vec{Q}_1^4}{(Q_1^0)^2}F_+^{(3)}(\mpc,\mpc,\mpn,\mpc,s_1)\nn
%(3)
&+ 2 K_0 C_{x}(\tilde{s}_1^{0+})\Bigl(C_{+0}(s_1)+E_{+0}^{-}(s_1,s_2,s_3)\Bigr) F_+(\mpn,\mpn,\mpn,\mpc,s_1) \nn
&\quad - 4K_0\Bigl[\meta D_{x}\Bigl(C_{+0}(s_1)+E_{+0}^{-}(s_1,s_2,s_3)\Bigr)\frac{\vec{Q}_1^2}{Q_1^0} + C_{x}(\tilde{s}_1^{0+})E_{+0}(s_1,s_2,s_3)\Bigr] \nn
&\qquad \times F_+^{(1)}(\mpn,\mpn,\mpn,\mpc,s_1) \nn
&\quad + 8\meta K_0 D_{x}E_{+0}(s_1,s_2,s_3)\frac{\vec{Q}_1^2}{Q_1^0}F_+^{(2)}(\mpn,\mpn,\mpn,\mpc,s_1) + (s_1 \leftrightarrow s_2) \biggr\}\nn
%(4)
& + 2 K_0 C_{00}(\tilde{s}_3^{00}) C_{x}(s_3)F_0(\mpn,\mpn,\mpn,\mpn,s_3)\nn
&\quad-4\meta K_0 D_{00} C_{x}(s_3) \frac{\vec{Q}_3^2}{Q_3^0} F_0^{(1)}(\mpn,\mpn,\mpn,\mpn,s_3)\nn
%(5)
& + 4 L''(s_3) C_x(\tilde{s}_3^{00}) C_{x}(s_3) F_0(\mpc,\mpc,\mpn,\mpn,s_3)\nn
&\quad + 4\Bigl[L_1 C_{x}(\tilde{s}_3^{00})-2\meta L''(s_3) D_x\Bigr] C_{x}(s_3) \frac{\vec{Q}_3^2}{Q_3^0} F_0^{(1)}(\mpc,\mpc,\mpn,\mpn,s_3)\nn
&\quad - 8\meta L_1 D_x C_{x}(s_3)\frac{\vec{Q}_3^4}{(Q_3^0)^2} F_0^{(2)}(\mpc,\mpc,\mpn,\mpn,s_3)\nn
%(6)
& + 8 \Bigl[L'''_{+0}(s_3)C_{+0}(\tilde{s}_3^{+-})\nn
&\qquad- \frac{\Delta_\pi}{4\mpn} \Bigl(\frac{s_3+2\vec{Q}_3^2}{2Q_3^0}-p_3^0\Bigr) L_0 E_{+0}\Bigr]C_{+-}(s_3) F_0(\mpc,\mpn,\mpc,\mpc,s_3) \nn
&\quad - 8\Bigl[\frac{1}{2}L_1 C_{+0}(\tilde{s}_3^{+-}) + 2\meta L_{+0}'''(s_3)D_{+0} \nn
&\qquad- \frac{\Delta_\pi}{4\mpn} L_0 E_{+0}\Bigr]C_{+-}(s_3)\frac{\vec{Q}_3^2}{Q_3^0} F_0^{(1)}(\mpc,\mpn,\mpc,\mpc,s_3) \nn
&\quad + 8\meta L_1 D_{+0}C_{+-}(s_3)\frac{\vec{Q}_3^4}{(Q_3^0)^2} F_0^{(2)}(\mpc,\mpn,\mpc,\mpc,s_3)~,  \label{eq:CA} \\[5mm]
\M_c^B
&=\Bigl\{4L'(s_1)C_{+0}^2(s_1)J_{+0}^2(s_1)+(s_1 \leftrightarrow s_2)\Bigr\}+K(s_3)C_{00}(s_3)C_x(s_3)J_{00}^2(s_3)\nn
&+2\Bigl[L(s_3)C_x^2(s_3)+K(s_3)C_x(s_3)C_{+-}(s_3)\Bigr]J_{+-}(s_3)J_{00}(s_3)+4L(s_3)C_{+-}^2(s_3)J_{+-}^2(s_3)~.\label{eq:CB} 
\end{align}
We have used the following abbreviations:
\begin{align}
J_{ab}(s_i)&=\frac{iq_{ab}(s_i)}{8\pi\sqrt{s_i}}~,\qquad q_{ab}^2(s_i)=\frac{\lambda(s_i,M_a^2,M_b^2)}{4s_i}~,\nn
C_n(s_i) &= C_n + D_n \big(s_i - s_n^{\rm thr}\big) + F_n \big(s_i - s_n^{\rm thr}\big) ~, \nn
\tilde s_i^{cd} &= M_c^2+M_i^2-s_i+\frac{\meta}{Q_i^0}\big(s_i+2\vec{Q}_i^2 -M_c^2+M_d^2 \big) ~,\nn
\tilde{E}_{+0}(s_1,&s_2,s_3)=E_{+0}\frac{q_{+0}^2(s_1)}{3s_1\meta}\Bigl(s_1(s_3-s_2)+\Delta_\pi(\mpc^2-\meta^2)\Bigr)~,\nn
E_{+0}^{(\pm)}(s_1,&s_2,s_3) = E_{+0} \biggl[\frac{(s_1(\pm\Delta_\pi))(s_3-s_2+\Delta_\pi)}{2\meta Q_1^0}-\Delta_\pi\biggr] ~,\nn
E_{+-}(s_1,&s_2,s_3) = E_{+-} \, \frac{s_3(s_1-s_2)}{2\meta Q_3^0} ~,\nn
K(s_i) &= K_0 + K_1 \Bigr[\left(p_i^0-\mpn\right)^2 + 2\left(\frac{Q_i^0}{2}-\mpn\right)^2+\frac{\vec{Q}_i^2}{6}\left(1-\frac{4\mpn^2}{s_i}\right)\Bigr]~,\nn
L(s_i) &= L_0 + L_1\left(p_i^0-\mpn\right)+L_2\left(p_i^0-\mpn\right)^2+L_3\frac{\vec{Q}_i^2}{3}\,
\Bigl(1- \frac{4\mpc^2}{s_i}\Bigr) \nn
\tilde{L}(s_i) & =L_1+2L_2\biggl[\frac{Q_i^0}{2}\Bigl(1-\frac{\Delta_\pi}{s_i}\Bigr)-\mpn\biggr]+2L_3\biggl[p_i^0-\frac{Q_i^0}{2}\Bigl(1+\frac{\Delta_\pi}{s_i}\Bigr)\biggr]\nn
L'(s_i) &= L_0 + L_1\Bigl(\frac{Q_i^0}{2}\Bigl(1-\frac{\Delta_\pi}{s_i}\Bigr)-\mpn\Bigr)+L_2\biggl[\Bigl(\frac{Q_i^0}{2}\Bigl(1-\frac{\Delta_\pi}{s_i}\Bigr)-\mpn\Bigr)^2+\frac{\vec{Q}_i^2}{3s_i}\,
q_{+0}^2(s_i)\biggr] \nn
& \quad + L_3\biggl[\Bigl(\frac{Q_i^0}{2}\Bigl(1+\frac{\Delta_\pi}{s_i}\Bigr)-p_i^0\Bigr)^2+\frac{\vec{Q}_i^2}{3s_i}\,
q_{+0}^2(s_i)\biggr] ~,\nn
L''\big(s_i^{(\pm)}\big) &= L_0 + L_1 \Big( \frac{s_i (\pm \Delta_\pi)}{2Q_i^0} -\mpn \Big)~, ~~ \nn
L'''_{ab}\big(s_i^{(\pm)}\big) &= L_0 + L_1 \Big( \frac{1}{2}(\meta-M_a-M_b) - \frac{s_i(\pm \Delta_\pi)}{4Q_i^0}\Big)~.
\end{align}
In the notation, it is understood that the shape parameter term $F_n$ is omitted in the 
polynomials $C_n(\tilde s_i^{cd})$ inside the ``genuine'' two-loop graphs. 
There is a subtlety with regard to the neutral Dalitz plot couplings in the irreducible two-loop graphs: since we included these couplings only up to $\order(\epsilon^2)$,
the $\Delta I=1$ rule is only fulfilled up that same order. Eq.~\eqref{eq:NREFTsym}, however, is valid up to $\order(\epsilon^4)$, so that
we have to replace $K_0\to\bar K_0=-(3L_0+L_1Q_n)$ in $\M_n^A$ and $\M_c^A$ above or simply $\bar{K}_0=\tilde{\N}_n(1-\frac{4}{9} \tilde{b} M_{\eta }^2 Q_n^2)$. 
The numerical effects of this replacement are small.

$F_i(\ldots;s)$, $F_i^{(k)}(\ldots;s)$, $k=1,\,2,\,3$,
stand for the integral representations $F(\ldots;s)$, $F^{(k)}(\ldots;s)$, 
evaluated at $\vec{Q}_i^2=\lambda(\meta^2,M_i^2,s_i)/4\meta^2$, with $i=1,2,3$.
The analytic expression for these two-loop functions read
\begin{align}
F&(M_a,M_b,M_c,M_d,s)   =   \N\,\Bigl[2A\, f_1+B\, f_0 -\frac{3\vec{Q}^2}{10s}(B\,f_1+2C\,f_0) \nn&\quad+ {\K
(X_3f_3+X_2f_2+X_1f_1+X_0f_0)\Bigr]}~, \nn
F^{(1)}&(M_a,M_b,M_c,M_d,s) = \frac{\N}{10}{(1+\delta)}\bigl[ (10A-B)f_1 + (5B-2C)f_0 \bigr] + \Order(\epsilon^4)~, \nn
F^{(2)}&(M_a,M_b,M_c,M_d,s) = \frac{\N}{2} \biggl[ - \frac{1}{\vec{Q}^2} \Big(2A^2 f_3 + 3AB f_2 +(B^2-2AC) f_1 + BC f_0 \Big)  \nn
&\quad+ \frac{(1+\delta)^2}{4} \Big( A f_3 + (B-2A) f_2 + (4A-2B+C) f_1 +2(B-C) f_0 \Big) \biggr] +\Order(\epsilon^4) ~,\nn
F^{(3)}&(M_a,M_b,M_c,M_d,s) = \frac{\N(1+\delta)}{16}\biggl[\frac{3}{\vec{Q}^2}\Big(A^2 f_4+2A(B-4A)f_3\nn
&\qquad\quad +(2AC+B^2-12AB)f_2+ (2BC-8AC-4B^2)f_1+(C^2-4BC)f_0\Big)\nn
&\qquad -(1+\delta)^2\Big(Af_4+(B-3A)f_3+(3A-3B+C)f_2\nn
&\quad\qquad+ (3B-4A-3C)f_1+(3C-2B)f_0\Big)\biggr]+\Order(\epsilon^4)~, \label{eq:functionF1}
\end{align}
with
\begin{align}
f_2&=-\frac{1}{5A}\,(3Bf_1+Cf_0)\, ,\quad\quad
f_3=-\frac{1}{7A^2}\,\Bigl[3(AC-B^2)f_1-BCf_0\Bigr]\, ,\nn
f_4&=\frac{1}{9A^2}\Bigl[3(ABC+5BC-5B^3)f_1-5(B^2C-C^2)f_0\Bigr]\, ,\nn
X_0&=HBC-RC\, , \quad X_1=H(2AC+B^2)-R(2B-C)\, , \quad X_2=3HAB-R(3A-\frac{3}{2}\,B)\, ,\nn
X_3& =2HA^2+2AR\, , \quad
H=-\frac{3}{2}\,\biggl(1+\frac{{\bf Q}^2}{3s}\biggr)\, ,\quad
R=\frac{{\bf Q}^2Q_0^2}{2s}\,(1+\tilde\delta)^2\, .
\end{align}
and
\begin{align}
\N&=\frac{1}{256\pi^3\sqrt{s}}\, 
\frac{\lambda^{1/2}(s_0,M_a^2,M_b^2)}{s_0\sqrt{\Delta^2-\frac{(1+\tilde\delta)^2}{4}\,\vec{Q}^2}} ~, \nn
\K&=
\biggl[\frac{1}{2(\meta^2+M_c^2)-(M_a+M_b)^2-s_0} + \frac{1}{s_0-(M_a-M_b)^2} - \frac{2}{s_0} \biggr] 
\frac{\meta^2}{s_0-\meta^2-M_c^2} ~, \nn
f_0&=4\bigl(v_1+v_2-\bar v_{2}+h\bigr) ~,\nn
f_1&=
\frac{4}{3}\,\bigl(y_1(v_1-1)+y_2(v_2-1)-\bar y_{2}(\bar v_{2}-1) +h \bigr) ~,\nn
h&=\frac{1}{2}\ln\biggl(\frac{1+\vec{Q}^2/s}{1+\bar{\vec{Q}}^2/\bar s}\biggr)  ~,
\quad \bar{\vec{Q}}^2=\vec{Q}^2(\bar s) ~, \quad \bar s=(M_c+M_d)^2 ~, \nn
v_i&=\sqrt{-y_i}\,\arctan\frac{1}{\sqrt{-y_i}}  ~,\quad i=1,2 ~;\quad
\bar v_{2}=\sqrt{-\bar y_{2}}\,\arctan\frac{1}{\sqrt{-\bar y_{2}}} ~, \quad \nn
y_{1,2} &= \frac{-B \mp \sqrt{B^2-4AC}}{2A} ~,\quad \bar y_{2}=y_2(\bar s) ~,  \nn
A&= -\frac{\vec{Q}^2}{s}\,(M_c^2+\Delta^2) ~,\quad
B=q_0^2-\Delta^2+\frac{\vec{Q}^2}{s}\, M_c^2 ~,\quad
C=-q_0^2 ~,\nn
s_0&=\meta^2+M_c^2-2\meta\biggl( M_c^2+\frac{\vec{Q}^2(1+\tilde\delta)^2}{4}\biggr)^{1/2} ~, \quad
q_0^2 =\frac{\lambda(s,M_c^2,M_d^2)}{4s} ~, \nn
\Delta^2&=\frac{\lambda(\meta^2,M_c^2,(M_a+M_b)^2)}{4\meta^2} ~,\quad
\tilde\delta=\frac{M_c^2-M_d^2}{s} ~.\label{eq:functionF2}
\end{align}

\subsection{Resummed amplitudes}
\label{app:NREFTunit}
In order to estimate the effects of higher-order corrections we iterate the bubble diagrams and the external vertex of the non-trivial two-loop graph. A diagrammatic expression of this iteration is shown in Fig.~\ref{fig:iteratedbubbles}.
Here we show the results including isospin violation.
For the bubble chain a coupled-channel resummation can be performed analogously to Ref.~\cite{Kl4}. We obtain
\begin{align}
  \M_n^{\rm{u}}&(s_1,s_2,s_3)=\frac{2 L(s_1) C_x(s_1) J_{+-}(s_1) + K(s_1)\Bigl[C_{00}(s_1) J_{00}(s_1) - 2\chi(s_1)J_{+-}(s_1) J_{00}(s_1)\Bigl]}{1 - 2 C_{+-}(s_1) J_{+-}(s_1) - C_{00}(s_1) J_{00}(s_1) + 2 \chi(s_1)J_{+-}(s_1) J_{00}(s_1)}\nn
					      &\quad + (s_1\leftrightarrow s_2) + (s_1\leftrightarrow s_3)~,\nn
  \M_c^{\rm{u}}&(s_1,s_2,s_3)=\frac{2C_{+0}(s_1)J_{+0}(s_1)L'(s_1)}{1 - 2C_{+0}(s_1)J_{+0}(s_1)} -\frac{\tilde{E}'_{+0}(s_1,s_2,s_3)\tilde{L}(s_1)J_{+0}(s_1)}{1-E_{+0}(s_1)J_{+0}(s_1)}+ (s_1\leftrightarrow s_2)\nn
					      &\quad + \frac{2 L(s_3) \Bigl[C_{+-}(s_3)J_{+-}(s_3) - 2\chi (s_3)J_{+-}(s_3) J_{00}(s_3)\Bigr] + K(s_3) C_x(s_3) J_{00}(s_3)}{1 - 2 C_{+-}(s_3) J_{+-}(s_3) - C_{00}(s_3) J_{00}(s_3) + 2 \chi (s_3)J_{+-}(s_3)J_{00}(s_3)}~,
\end{align}
where 
\begin{align}
\chi(s_i)&=C_{+-}(s_i) C_{00}(s_i)-C_x(s_i)^2~,\nn 
\tilde{E}'_{+0}(s_1,s_2,s_3)&=\Bigr[E_{+0}+G_{+0}(s_i-s_{+0}^{\text{thr}})\Bigl]\frac{q_{+0}^2(s_1)}{3s_1\meta}\Bigl(s_1(s_3-s_2)+\Delta_\pi(\mpc^2-\meta^2)\Bigr)~,\nn
E_{+0}(s_i)&=\frac{4q_{+0}^2(s_i)}{3}\Bigr[E_{+0}+G_{+0}(s_i-s_{+0}^{\text{thr}})\Bigl]~,\qquad~G_{+0}=3\pi  b_1~,  
\end{align}
and $b_1$ is the P-wave effective range. Additionally, we performed a resummation of the external vertex of the non-trivial two-loop diagram. This can be achieved by replacing the outer vertex in Eqs.~\eqref{eq:NB} and~\eqref{eq:CB} according to
\begin{align}
C_{00}(s_i)&\to\frac{ C_{00}(s_i)-2\chi(s_i)J_{+-}(s_i) }{1-C_{00}(s_i)J_{00}(s_i)-2C_{+-}(s_i)J_{+-}(s_i) +2\chi(s_i)J_{+-}(s_i) J_{00}(s_i)}~,\nn
C_{x}(s_i)&\to\frac{ C_{x}(s_i) }{1-C_{00}(s_i)J_{00}(s_i)-2C_{+-}(s_i)J_{+-}(s_i) +2\chi(s_i)J_{+-}(s_i) J_{00}(s_i)}~,\nn
C_{+0}(s_i)&\to\frac{C_{+0}(s_i)}{1-2C_{+0}(s_i)J_{+0}(s_i)}~,\qquad E_{+0}^{(\pm)}(s_1,s_2,s_3)\to\frac{ {E'}_{+0}^{(\pm)}(s_1,s_2,s_3) }{1-E_{+0}(s_1)J_{+0}(s_1)}~,\nn
C_{+-}(s_i)&\to\frac{C_{+-}(s_i)-\chi(s_i)J_{00}(s_i)}{1-C_{00}(s_i)J_{00}(s_i)-2C_{+-}(s_i)J_{+-}(s_i)+2\chi(s_i)J_{+-}(s_i)  J_{00}(s_i)}~, 
\end{align}
with
\be
 {E'}_{+0}^{(\pm)}(s_1,s_2,s_3) = \Bigr[E_{+0}+G_{+0}(s_i-s_{+0}^{\text{thr}})\Bigl] \biggl[\frac{(s_1(\pm\Delta_\pi))(s_3-s_2+\Delta_\pi)}{2\meta Q_1^0}-\Delta_\pi\biggr] ~.
\ee

\section{Comment on imaginary parts of two-loop diagrams}\label{app:3cut}

\FIGURE[t]{
 \centering
 \includegraphics[width=0.6\linewidth]{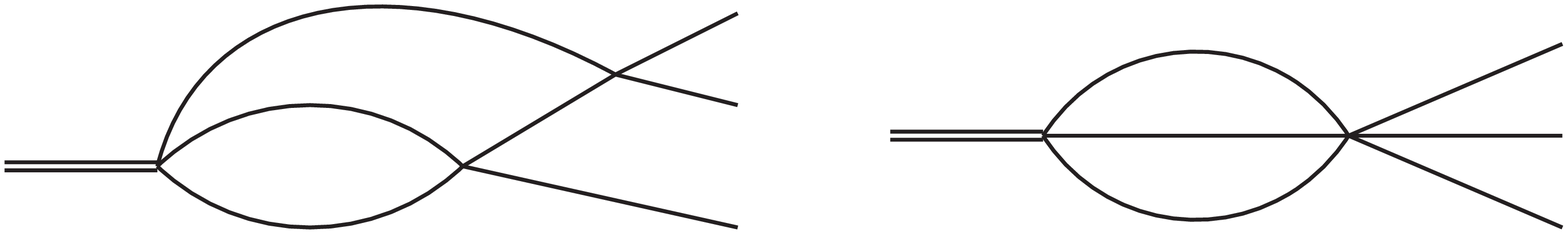}
 \caption{Two-loop graphs with three-particle cuts in the physical region.
 \label{fig:3cut}}
}

In our analysis of the non-relativistic $\eta\to3\pi$ decay amplitude, 
we have neglected the imaginary parts of the non-trivial two-loop graphs, 
see Fig.~\ref{fig:3cut}~(left).  The loop function $F(s)$ (in the simplified notation
introduced for the equal-mass case in Sect.~\ref{sec:isospinlimitresults}) 
given in Appendix~\ref{app:NREFT2loop} strictly speaking only corresponds to the \emph{real}
part of this diagram.  At leading order in the $\epsilon$ expansion, its \emph{imaginary} part
is given by
\be
{\rm Im}\,F(s) = -\frac{1}{\sqrt{3}(32\pi)^2} \frac{(M_\eta-3M_\pi)^2}{M_\pi^2} + \Order(\epsilon^6) ~.
\label{eq:ImF}
\ee
We therefore confirm that ${\rm Im}\,F(s) = \Order(\epsilon^4)$, while the real part of the same
diagram already starts at $\Order(\epsilon^2)$.  The imaginary part is due to the three-pion cut
and only arises because the $\eta$
is unstable, $M_\eta > 3M_\pi$. It stems from a part of the non-relativistic loop integral in which
one of the propagators is non-singular, and therefore yields a result very similar to that 
of the sunset graph Fig.~\ref{fig:3cut}~(right), which in the non-relativistic framework can only
arise when introducing (very small) six-pion vertices~\cite{K3pi,GKR}.  
The three-pion cut causes a non-vanishing imaginary part of the isospin amplitudes $\mathcal{M}_I(s)$
already at threshold $s=4M_\pi^2$. 

It is obvious that ${\rm Im}\,F(s)$ can only contribute to the amplitude's squared modulus 
at $\Order(a_{\pi\pi}^3\epsilon^5)$ via interference with one-loop terms, and is therefore naturally
suppressed compared to the real part at two-loop order.  
What is less clear is its relative importance compared to the imaginary parts generated
at three loops by the unitarization prescription Eq.~\eqref{eq:unitarized}. The latter contributes
to $|\mathcal{M}|^2$ at $\Order(a_{\pi\pi}^4\epsilon^4)$, i.e.\ it is suppressed in powers of $a_{\pi\pi}$, 
but enhanced in $\epsilon$.  
By investigating the imaginary part of the dominant isospin $I=0$ amplitude $\mathcal{M}_0(s)$, 
with Eq.~\eqref{eq:ImF} added appropriately to the representation Eq.~\eqref{eq:isoamps}, 
we find that the two-loop imaginary part is suppressed by more than a factor of 30 
relative to the one-loop piece at the center of the Dalitz plot, and by roughly a factor of 2
relative to the three-loop part.  This suppression grows even stronger when considering
derivatives of the amplitude, as $s$-dependence in ${\rm Im}\,F(s)$ is even further
suppressed in $\epsilon$, see Eq.~\eqref{eq:ImF}.
We therefore neglect these terms of $\Order(i a_{\pi\pi}^2 \epsilon^4)$ in our analysis, and 
consider their effects to be safely included in our error estimates due to partial higher-order
resummation.
The smallness of the imaginary parts due to three-pion cuts is in accordance with 
findings from ChPT at two loops~\cite{BGeta3pi} as well as from dispersion relations~\cite{CLPeta3pi}.

\end{document}